\documentclass[journal,twocolumn]{IEEEtran}

\usepackage{amsmath,graphicx}
\usepackage{cite}
\usepackage[normalem]{ulem}
\DeclareMathOperator{\E}{\mathbb{E}}
\usepackage{amsmath,amssymb,amsfonts}

\usepackage{graphicx}
\usepackage{textcomp}
\usepackage{xcolor}
\usepackage{bm}
\newtheorem{exmp}{Example}[section]
\usepackage{subcaption}
\usepackage{algorithm}
\usepackage[noend]{algpseudocode}
\usepackage{epsfig}
\usepackage{float}
\usepackage{comment}
\usepackage{mathtools,amssymb}
\usepackage{needspace}
\usepackage[T1]{fontenc}
\usepackage[utf8]{inputenc}
\usepackage{pgfplots}
\usetikzlibrary{arrows}
\usetikzlibrary{datavisualization.formats.functions}
\usepackage{grffile}
\pgfplotsset{compat=newest}
\newtheorem{assumption}{Assumption}
\newtheorem{mydef}{Definition}

\newtheorem{proof}{Proof}
\newtheorem{proposition}{Proposition}
\usepackage{tikz}
\newcommand{\norm}[1]{\left\lVert#1\right\rVert}
\usetikzlibrary{plotmarks}
\usetikzlibrary{arrows.meta}
\usepackage{diagbox}
\usepgfplotslibrary{patchplots}
\usepackage{amsmath}
\usepackage[utf8]{inputenc}
\usepackage[english]{babel}

\newcommand{\RNum}[1]{\uppercase\expandafter{\romannumeral #1\relax}}

\DeclareMathOperator*{\argmax}{arg\,max}
   
\def\BibTeX{{\rm B\kern-.05em{\sc i\kern-.025em b}\kern-.08em
		T\kern-.1667em\lower.7ex\hbox{E}\kern-.125emX}}
\begin{document}
	
	\title{A Reinforcement Learning based approach for Multi-target Detection in Massive MIMO radar\\
		\thanks{Funded by the Deutsche Forschungsgemeinschaft (DFG,
			German Research Foundation) Project-ID 287022738 TRR
			196 (S03).}
	 \thanks{Profs. Gini's and Greco's work supported by the Italian Ministry of Education and Research (MIUR) within the framework of the CrossLab Project (Departments of Excellence program).}}
	
	\author{Aya Mostafa Ahmed,~\IEEEmembership{Student Member,~IEEE,} Alaa Alameer Ahmad,~\IEEEmembership{Student Member,~IEEE,} \\Stefano Fortunati,~\IEEEmembership{Senior Member,~IEEE,} Aydin Sezgin,~\IEEEmembership{Senior Member,~IEEE,} Maria S. Greco,~\IEEEmembership{Fellow,~IEEE,} Fulvio Gini,~\IEEEmembership{Fellow,~IEEE}
	\thanks{
		A. M. Ahmed, A. A. Ahmad, and A. Sezgin are with the Institute of Digital
		Communication Systems, Ruhr University Bochum, 44801 Bochum, Germany
		(e-mail: aya.mostafaibrahimahmad@rub.de; alaa.alameerahmad@rub.de;
		aydin.sezgin@rub.de).}
	\thanks{S.~Fortunati, is with Universit\'{e} Paris-Saclay, CNRS, CentraleSup\'{e}lec, Laboratoire des signaux et syst\'{e}mes, 91190, Gif-sur-Yvette, France \& DR2I-IPSA, 94200, Ivry sur Seine, France. (e-mail: stefano.fortunati@centralesupelec.fr).}
\thanks{M.S.~Greco, and F. Gini is with Universit\`a di Pisa, Dipartimento di Ingegneria dell'Informazione, Via Caruso, 56122, Pisa, Italy. (e-mail:m.greco@iet.unipi.it; f.gini@ing.unipi.it).}}
	
	\maketitle
	
	\begin{abstract}
	This paper considers the problem of multi-target detection for massive multiple input multiple output (MMIMO) cognitive radar (CR). The concept of CR is based on the perception-action cycle that senses and intelligently adapts to the dynamic environment in order to optimally satisfy a specific mission. However, this usually requires a priori knowledge of the environmental model, which is not available in most cases. We propose a reinforcement learning (RL) based algorithm for cognitive multi-target detection in the presence of unknown disturbance statistics. The radar acts as an agent that continuously senses the unknown environment (i.e., targets and disturbance) and consequently optimizes transmitted waveforms in order to maximize the probability of detection ($P_\mathsf{D}$) by focusing the energy in specific range-angle cells (i.e., beamforming). \\ Furthermore, we propose a solution to the beamforming optimization problem with less complexity than the existing methods. Numerical simulations are performed to assess the performance of the proposed RL-based algorithm in both stationary and dynamic environments. The RL based beamforming is compared to the conventional omnidirectional approach with equal power allocation and to adaptive beamforming with no RL. As highlighted by the proposed numerical results, our RL-based beamformer outperforms both approaches in terms of target detection performance. The performance improvement is even particularly remarkable under environmentally harsh conditions such as low SNR, heavy-tailed disturbance and rapidly changing scenarios.
		
	\end{abstract}
	
	
	\section{Introduction}
Cognitive radar (CR) is described as a radar system which senses the environment, learns from it and makes decisions based on what it has learned to accomplish certain assigned tasks though a perception-action cycle of cognition. 
	In a non-stationary environment, this cycle is repeated continuously. The non-stationarity can be caused by statistical weather/sea variations, stochastic disturbance (i.e., Gaussian noise plus clutter) or the presence of unknown non-static targets. \\
	In \cite{Haykin2}, Haykin clearly distinguishes between traditional feed forward radar, fully adaptive radar and a CR. A radar is considered adaptive, when it employs a global feedback including the environment in this feedback loop, where adaptive filtering at the receiver or adaptive beamforming at the transmitter might be applied~\cite{adaptiveRadar}. Unlike the classical adaptive paradigm, CR develops its own behavior rules from the experience gained, stores it in the memory and extends this knowledge to the transmitter. This is followed by a set of smart decision-actions. Consequently, CRs would benefit significantly from the waveform diversity at the transmitter offered by multiple input multiple output (MIMO) radar systems.
	\\
	Unlike phased arrays, MIMO exploits multiple correlated or uncorrelated probing signals, offering higher degrees of freedom (DoF). There are two main types of MIMO radar systems: \textit{widely separated} and \textit{colocated}. A widely separated MIMO radar exploits the spatial diversity of the target's radar cross section (RCS) by using widely separated transmit/receive antennas\cite{WSradar}. On the other hand, the antennas of a colocated MIMO radar are closely spaced, allowing for significant coherent gain when combining the probing signals, which can be achieved through designing the transmit beampattern~\cite{colocRadar}. This latter type is quite appealing for CR systems, since the transmitter can optimize the beampattern based on the received radar echoes.
	\\
	Specifically, in this paper, we focus on colocated \textit{massive} MIMO (MMIMO). The MMIMO radar has been recently shown to be able to provide robustness against the unknown disturbance distribution \cite{fortunati2019massive}. 
	Moreover, as foreseen in \cite{massivemimosurvey,drone}, thanks to its higher DoFs, a MMIMO can detect small unmanned vehicles (UAV) whose RCS can be up to three orders of magnitude smaller than manned vehicles. 
	However, together with the benefits, the MMIMO paradigm brings with it new challenging issues. One of the main open problems of MMIMO radar is the design of robust algorithms for detection and estimation with scalable complexity as the number of deployed antennas increases. Moreover, cognitive MMIMO radar, requires optimizable waveforms \cite{greco2018cognitive}. Hence, the design of scalable and fast accurate optimization algorithms is necessary. The current existing work relying on semi-definite programming (SDP) to solve for the beampattern optimization problem would impose further computational complexity and processing overhead to the MMIMO problem. This is due to the fact that in those papers, the beamforming is done over two computationally demanding steps. In the first step, the covariance matrix is optimized. In the second step, this matrix is used to synthesize the beamformers.\\
	Complementary and equally essential as the waveform diversity at the transmitter, is the receiver cognition, which guides the radar decisions and controls the choice of the waveforms. In~\cite{Haykin,CRBell,Haykin2}, the authors utilized Bayesian filtering for the perception-action cycle, where the receiver makes probabilistic predictions on the next environmental state given the current state. This approach however may lead to model mismatches as it depends on prior information of the environment dynamics. To avoid this dependence, reinforcement learning (RL) is adopted for cognitive radars in \cite{CRRL1,CRRL2}. \\RL is a learning approach addressing model free problems by using software defined agents, which learn from the observations collected from the environment, and take the best possible actions according to a reward function \cite{Sutton1998}. Those interactions with the environment are formally described as a Markov decision process (MDP). 
	\\In this paper, we propose an RL-based multi-target detection algorithm for a MMIMO CR systems. The agent in our setup is the radar and the environment contains targets and disturbance. We assume no prior information about the environment, in particular, no assumptions will be done on the the number of targets and on the statistical model of the disturbance.
	\subsection{Related Work} 
	 The use of machine learning with CR has been recently explored in the literature. In \cite{CRRL1,CRRL2} RL-CR is used for dynamic spectrum allocation, while in \cite{Simeone} the authors use CR for target detection in an end-to-end learning approach. They propose an alternating procedure to jointly design the transmit waveform and the detector. A neural network is used to approximate the generalized likelihood ratio test (GLRT) while the transmit waveform is fixed. Consequently, for a fixed detector, they train the transmitted waveform using deep RL. However, there is no statistical guarantee on the resulting detection performance that may seriously degrade in the presence of a mismatch between the actual disturbance model and the one used for training. Moreover, they don't address multi-target detection.\\ In \cite{MLCGR}, the authors use machine learning approaches to estimate the optimal detection threshold, based on non-linear transformation of the order statistics. They use an offline library for the disturbance distributions, where they assume a priori known covariance matrix to build this library. \\In \cite{RLUAV}, RL has been used in indoor mapping for UAV applications. Mechanical beamforming using the UAV rotation is used for target detection. However, only Gaussian noise is considered in the detection algorithm. \\ In the context of MMIMO, the work in \cite{massivemimo_nongaussian} exploits random matrix theory to analyses the detection performance. However, the approach proposed in \cite{massivemimo_nongaussian} requires the cumbersome requirement of a large number of observations. \\Unlike the previous approaches, we propose a fully data-driven algorithm that does not require any offline knowledge of possible disturbance distributions nor their covariance matrix and thus avoids issues with model-mismatches.	
	\subsection{Contributions}
This paper includes several contributions which can be summarized as the following :
	\begin{enumerate}
		\item We derive an original RL-based MMIMO CR detection algorithm that does not rely on any prior
		information about statistical model of the disturbance, or
		the number of targets.
		\item By exploiting the specific feature of a MMIMO system described in \cite[Corollary 1]{fortunati2019massive}, we propose a reward function calculated in terms of the probability of detection ($P_\mathsf{D}$) regardless of the disturbance distribution.
		\item We propose a beamforming optimization approach that can scale up with a large number of antenna, where we optimize directly for the beamforming matrix without the need to optimize for the covariance matrix first.
	
		\item As suggested by the numerical results, the algorithm is able to detect low SNR targets with a radar operating under a constraint on the probability of false alarm $P_{\mathsf{FA}}=$ $10^{-5}$. Moreover, it is robust to environmental changes, e.g., it can detect fading targets and targets changing their angular positions.
	\end{enumerate}
\textbf{Notations}: In this paper we used upper case letters $\mathbf{ A }$ and lower case ones $\mathbf{ a}$ to denote matrices and vectors respectively. $\left(\cdot\right)^T$ and $\left(\cdot\right)^H$ denote a matrix transpose and conjugate transpose respectively, while $a^*$ denotes the conjugate of $a$. $\mathbf{I}_N$ denotes identity matrix of size $N \times N$, while $\E[\cdot]$ denotes the statistical expectation. The Kronecker product is represented by $\otimes$. A closed interval of numbers between $a$ and $b$ is denoted by $\left[a,b\right]$, while a set containing only $a$ and $b$ is denoted by $\{a,b\}$. The absolute value is represented by $\left |\cdot\right |$ and the real value is denoted by $\rm{Re}\{\cdot\}$.
	\section{Problem Formulation}
	\label{PROBLEM}
We consider a colocated MIMO radar system with $N_T$ transmit antennas and $N_R$ receiver antennas. Both are uniform linear arrays (ULA) with $\frac{\lambda}{2}$ spacing between the antennas, where $\lambda$ is the operating wavelength. 
	\subsection{System Model}
 The complex baseband version of the received signal at continuous time $t$ reflected from one point-like target is defined as \cite{TXBEAMmodel,colocRadar}
	\begin{align}
	\mathbf{\hat{y}}(t) & = \alpha \mathbf{a}_R(\theta) \mathbf{a}_T^T(\theta)\mathbf{s}(t-\tau)+\mathbf{c}(t)
	\label{recievedsigY}
	\end{align}
	where $\mathbf{\hat{y}}(t) \in \mathbb{C}^{N_R}$. The transmit and receive arrays are characterized by the array manifolds: $\mathbf{a}_T(\theta)$ and $\mathbf{a}_R(\theta)$, respectively, where $\theta$ is the target direction. Hence, $\mathbf{a}_R(\theta) = [1,e^{j\frac{2\pi d}{\lambda} sin\theta},\dots,e^{j\frac{2\pi d}{\lambda} (N_R-1)sin\theta}]^T$ and $\mathbf{a}_T(\theta)$ is defined similarly. $\alpha \in \mathbb{C}$ is a deterministic unknown variable which accounts for the radar RCS and the two way path loss following Swerling 0, while $\tau$ is the time delay due to the target position with respect to the radar. $\mathbf{c}(t) \in \mathbb{C}^{N_R} $ is the random disturbance vector, which is produced by clutter and white Gaussian noise. $\mathbf{s}(t)  \in \mathbb{C}^{N_T}$ is the transmit signal from all $N_T$ antennas, generated as linear combination of independent orthonormal signals $\mathbf\Phi (t)\in \mathbb{C}^{N_T}$, where
	\begin{align}
	\textbf{s}(t) & = \mathbf{W} \mathbf\Phi (t),
	\label{txSignal}
	\end{align}
	and $\mathbf{W} = [\mathbf{w}_1,\dots,\mathbf{w}_{N_T}]^T \in \mathbb{C}^{N_T \times {N_T}} $, $\mathbf{w}_m \in \mathbb{C}^{N_T}$ describe the beamforming weight matrix. Moreover, $\mathbf{W}$ is a square matrix which must obey the trace constraint $\mathrm{tr}\{\mathbf{W}\mathbf{W}^H\}=P_T$, where $P_T$ is the total transmit power.
	Furthermore, the beam pattern produced by the transmitted waveforms can be expressed as $B(\theta) = \mathbf{a}_T^T(\theta)\mathbf{R}_W\mathbf{a}_T^*(\theta)$, where $\mathbf{R}_W=\mathbf{W}\mathbf{W}^H$ \cite{colocRadar,TXBEAMmodel}.\\
	At the receiver, the received signal is processed by a linear matched filter $\mathbf\Phi(t)$ tuned at delay $\hat{\tau}$ considering a single transmitted pulse such that
	\begin{align}
	\mathbf{Y}(\hat{\tau})=\int_{0}^{T} \mathbf{\hat{y}}(t)\mathbf\Phi^{H} (t-\hat{\tau}) dt.
	\end{align}
	Hence, 
	\begin{align}
    \textbf{Y} & = \alpha \mathbf{a}_R(\theta)\mathbf{a}_T^T(\theta)\mathbf{W}\int_{0}^{T}\mathbf\Phi (t-{\tau})\mathbf\Phi^{H} (t-\hat{\tau}) dt+\textbf{C},
    \label{recievedsigY2}
\end{align}
where $ \textbf{Y} \in \mathbb{C}^{N_R\times N_T}$ and $\textbf{C}= \int_{0}^{T} \mathbf{c}(t)\mathbf\Phi^{H} (t-\hat{\tau}) dt.$
We assume that the matched filter is perfectly tuned to the target delay\cite{TXBEAMmodel}, hence $\hat{\tau}=\tau$ and $\int_{0}^{T}\mathbf\Phi (t-{\tau})\mathbf\Phi^{H} (t-\hat{\tau}) dt=\mathbf{I}$.
Rewriting eq. \eqref{recievedsigY} to be in a vector form as
	\begin{align}
\mathbb{C}^{N_RN_T} \ni	\textbf{y} & = \text{vec}(\textbf{Y}) = \alpha \textbf{h}(\theta)+\textbf{c} ,
	\label{received_signal}
	\end{align}
	where $\text{vec}(\cdot)$ denotes the vectorization operator. \\
	Moreover $\textbf{c}=\text{vec}(\textbf{C})$ denotes the spatially colored disturbance vector. Then, utilizing the properties of the Kronecker product, the vector \textbf{h} is defined as:
	\begin{align}
	\textbf{h}(\theta) & = (\textbf{W}^T \mathbf{a}_T (\theta)) \otimes \mathbf{a}_R (\theta).
	\label{channel}
	\end{align}
	It is worth mentioning that the sampling procedure (in fast-time, i.e. in range) is assumed to be done correctly respecting Nyquist theorem after the matched filter \cite{TXBEAMmodel}.
	\subsection{Disturbance Model}
		The statistical characterization of disturbance is a difficult task \cite{massivemimosurvey}, generally, it is usually unknown. Despite the fact that various disturbance models have been adopted in radar literature, the dynamic nature of the clutter may limit their validity \cite{Stefano_misspecified}. To avoid any model misspecification, robust approaches should be used.\\
		Here, building upon the results obtained in \cite{fortunati2019massive}, we adopt the following, very general, disturbance model:
	\begin{assumption}\cite{fortunati2019massive}
		\label{assumption}
	\textit{Let $\{c_n$: $\forall$ $n\}$ be the true and thus unknown disturbance process, which is a stationary discrete and circular complex valued process. It is only assumed that its autocorrelation function $r_C[m] \overset{\Delta}{=} \E\{c_n c_{n-m}^*\}$ has a polynomial decay}. 
	\end{assumption}
	It should be noted that such assumption weak enough to encompass all the most practical disturbance models such as autoregressive (AR), ARMA or general correlated non-Gaussian models \cite{fortunati2019massive}. \\  
	 In order to detect the presence of targets embedded in unknown disturbance, the detection problem is formulated as binary hypothesis testing problem described in the next subsection.
	
	\subsection{Detection Problem}
	It is assumed that the received signal in \eqref{received_signal} is processed by a bank of spatial filters, each tuned to a specific angle range. Such filter divides the radar field of view into separate $L$ discrete angle bins, such that $ \left\{\theta_l;l=1,\hdots,L\right\}$, and in total the system transmits $K$ pulses, $k\in \left\{1,\dots,K\right\}$. Hence, after spatial filtering, for angle bin $l$, at time $k$, eq. \eqref{received_signal} can be redefined as
		 \begin{align}
			\textbf{y}^k_{l} & =\mathbf{\alpha}^k_{l} \textbf{h}^k_l + \textbf{c}^k_{l}  .
		 \end{align} 
 The hypothesis testing problem for each angle bin $l$ is casted as 
	\begin{align}
	\label{HT}
	H_0:       & \quad \textbf{y}^k_{l} = \textbf{c}^k_{l} \hspace{0.2in} k=1,\dots,K\\
	H_1:       & \quad \textbf{y}^k_{l} =\mathbf{\alpha}^k_{l} \textbf{h}^k_l + \textbf{c}^k_{l} \hspace{0.2in} k=1,\dots,K.\nonumber
	\end{align}
The null hypothesis $H_0$  indicates that the cell under test contains only disturbance, i.e., clutter and noise, while the  alternative $H_1$ denotes single target detection. The entries of $\textbf{c}_l^k$ are sampled from complex random process, satisfying the general Assumption 1, having unknown covariance matrix $\mathbf{\Gamma}=\E\{(\textbf{c}_l^k)(\textbf{c}_l^k)^H\}$. The detection is performed per pulse, as the targets number, corresponding spatial angles and SNR can change from pulse $k$ to another. This also applies on the disturbance statistics, it can change in time and space. Hence, we assume a single snapshot scenario. In order to differentiate between $H_0$ and  $H_1$ in \eqref{HT}, a test statistic is required, where
	\begin{equation}
	\Lambda\left(\textbf{y}^k_{l}\right)\underset{H_0}{\overset{H1}{\gtrless}} \lambda.
	\label{thresh1}
	\end{equation}
	Since in radar applications it is of fundamental importance to control the $P_{\mathsf{FA}}$, the threshold $\lambda$ has to be chosen chosen to satisfy the following
	\begin{equation}
 \text{Pr}\{\Lambda\left(\textbf{y}^k_{l}\right)>\lambda |H_0\}= \int\limits_\lambda^\infty p_{{\Lambda}|H_0}(a|H_0) da = P_{\mathrm{FA}},
 \label{threshold}
	\end{equation}
	where $p_{\Lambda|H_0}$ is the probability density function (pdf) of $\Lambda\left(\textbf{y}^k_{l}\right)$ under $H_0$.
		 Usually conventional model based test statistics as \textit{generalized likelihood ratio test (GLRT)}, or \textit{Wald test} are used to solve for \eqref{thresh1}, yet they can not be directly applied here. This is due to the fact that the functional form of the pdf of $\textbf{c}^k_l$ is unknown. Instead, we apply a robust Wald-type detector derived in \cite{fortunati2019massive}, which requires the disturbance model only to satisfy Assumption 1. This detector is asymptotically distributed, (i.e., $N\rightarrow \infty$) as chi-squared $\chi^{2}_{2}$ random variable under both $H_0$ and  $H_1$ (see Appendix A) and, for each angle bin l, it is defined as follows:
	
	\begin{equation}
\Lambda^k_{l,\mathsf{RW}} = \frac{2|(\mathbf{h}^k_l)^H \mathbf{y}^k_l|^2}{(\mathbf{h}^k_l)^H \widehat{\mathbf{\Gamma}}\mathbf{h}^k_l},
\label{lambda_k}
	\end{equation}
where $\widehat{\mathbf{\Gamma}}$ is the estimate of the unknown $\mathbf{\Gamma}$. Further details about the calculation of 
		$\widehat{\mathbf{\Gamma}}$ and the asymptotic distribution of $ \Lambda^k_{l,\mathsf{RW}}$ are provided in Appendix A. \\The threshold $\lambda$ which satisfies eq. \eqref{threshold} regardless the disturbance distribution, can be expressed as \cite{fortunati2019massive}:
%
%
\\	\begin{equation}
\lambda = H^{-1}_{\chi ^2_2}(1-P_{\mathsf{FA}}),
\label{PFA}
\end{equation} 
in which $H^{-1}_{\chi ^2_2}(\cdot)$ is the inverse of the CDF function  $H_{\chi ^2_2}(\lambda) =~ \int\limits_{-\infty}^\lambda p_{\Lambda^k_{l,\mathsf{RW}}}(a|H_0)da$.\\
Eq. \eqref{PFA} guarantees the constant false alarm rate (CFAR) property for the robust Wald-type detector in \eqref{lambda_k}. However, the CFARness is only one aspect of a radar detection scheme. The second primary goal is to have good detection performance in terms of $P_\mathsf{D}$. In the next section, we propose a RL based algorithm that is able to enhance the detection while maintaining the CFAR property.
   
	\section{RL-BASED MMIMO COGNITIVE RADAR}
	In this section, we start by explaining the radar interpretation of the basic building blocks of the RL theory. Then, a detailed description of the proposed RL-based multi-target detection algorithm is provided.
	\subsection{Reinforcement Learning: a recall}
	RL is an area of machine learning, where an \textit{agent} learns how to make decisions to achieve a certain goal. This is done by trial and error interactions with the environment \cite{Sutton1998}. Typically an agent performs course of actions, then it evaluates its goal achievement through two types of information collected from the environment in response to those actions: its current \textit{state} and a \textit{reward}. The reward is defined as a scalar feedback signal, which the agent always seeks to maximize. It is specific to a certain task and a corresponding goal \cite{mlFOUNDATION}. 
The interactions with the environment in RL is formally described by Markov decision processes (MDP)\cite{Sutton1998}.
\begin{mydef}\textit{A Markov decision processes (MDP) is defined by a tuple \{$ \mathcal{S}$,$\mathcal{A}$,$\mathcal{P}$,${r}$\}, where $\mathcal{S}$ is a finite set of states, $\mathcal{A}$ is a finite set of actions, $\mathcal{P}$ is the transition probability from state ${s}$ to ${s}'$ $\in S$ after action $a \in A$ is performed, $r$ is the immediate reward evaluated after $a$ is executed.}
\end{mydef}
A \textit{policy} $\pi$: $\mathcal{S}\rightarrow \mathcal{A}$ is a function that maps a state $s\in \mathcal{S}$ into an action $a \in \mathcal{A}$. Moreover it defines which action has to be executed at each state. Thus, at time $k \in \left[0,K\right]$, the agent observes the state $s_k$, then based on a specific policy $\pi$, it takes action $a_k=\pi(s_k)$ . Consequently, a new state $s_{k+1}$ will be reached with probability $\mathcal{P}(s_{k+1}|s_k,a_k)$ and a reward $r_{k+1} \in \mathbb{ R }$ will be received. The observed information from the environment, the reward $r_{k+1}$ and $s_{k+1}$ are used to improve the policy. This process is repeated till the optimal policy is reached. \\ To provide a score to a given state, a \textit{state value function} $ \mathcal{V}_\pi : \mathcal{S}\rightarrow \mathbb{ R }$ is introduced. This function is defined as the expected cumulative reward received by the agent for starting from state $s$ and following policy $\pi$. More formally, the  \textit{state value function} for policy $\pi$ is defined as
\begin{equation}
\mathcal{V}_{\pi}(s)=\E_{\pi}\left[\sum_{k=0}^{\infty}\gamma ^k r_{k+1}|S_k=s\right],
\end{equation}
where $\E_{\pi}\left[\cdot\right]$ denotes the expected value of a random variable when the agent follows the policy $\pi$ at any time $k$. The scalar $\gamma \in \left[0,1\right]$  denotes the discount factor which controls the weight given to future rewards. 
In addition, let us define as $Q$-function the optimal \textit{action value function} for policy $\pi$ , where  \textit{$Q: \mathcal{S}\times\mathcal{A}\rightarrow \mathbb{ R }$} is defined as the expected cumulative reward for starting from state $s$ and taking action $a$ :
 \begin{equation}
 Q_{\pi}(s,a)=\E_{\pi}\left[\sum_{k=0}^{\infty}\gamma ^k r_{k+1}|s_k=s,a_k=a\right],
 \end{equation}
therefore the optimal \textit{state value function} can be written as 
 \begin{equation}
 \mathcal{V}^*_{\pi}(s)=\underset{a\in \mathcal{A}}{\arg\max} \hspace{0.1in} Q_{\pi}(s,a).
  \end{equation}
In the next subsection will map the aforementioned RL tools into our MMIMO radar setup.

		\subsection{SARSA algorithm and target detection}
	 The acronym SARSA is derived from the \textit{state-action-reward-state-action} sequence to update the values of the $Q$-function \cite{SARSA_expl}. SARSA is an on-policy RL algorithm, which evaluates and improves the same policy that is being used for action selection.  In contrast, off-policy algorithms, evaluate and improve a policy different  from the one used for action selection. SARSA falls under the category of model-free reinforcement learning algorithms, because it does not require a model of the environment.\\ In our radar problem, the agent has to maintain a state-action matrix \textbf{Q} $\in \mathbb{ R }^{(M+1) \times (M+1)}$ of elements $Q(s_k,a_k)$, where $M$ is the maximum number of targets that any MIMO radar can identify. This matrix is initialized with 0, afterwards based on the execution of a certain action, the agent shifts from one state to another, then updates the \textit{Q-function} according to the following update rule \cite{Sutton1998}
	\begin{align}
	\label{Qfun}
	Q\left(s_k,a_k\right)\leftarrow & Q\left(s_k,a_k\right)+\\& \alpha \left(r_{k+1}+ \gamma Q\left(s_{k+1},a_{k+1}\right) - Q\left(s_k,a_k\right)\right) \nonumber
	\end{align}
	The learning rate $\alpha \in \left[0,1\right]$ is used to control how much the recent experiences override the old ones. For instance, as $\alpha$ increases, the influence of the recent experiences on the \textit{Q function} increases. In the following subsections, the SARSA terms are explained according to the radar definitions.
	\subsection{The set of states}
	To define the state space $\mathcal{S}$, the statistic $ \Lambda^k_{l,\mathsf{RW}} $ from eq. \eqref{lambda_k} is utilized. If $ \Lambda^k_{l,\mathsf{RW}} $ is greater than the defined threshold ${\lambda}$ from eq. \eqref{PFA} for the angle bin $l$ at time $k$, a new statistic ${\bar{\Lambda}^k_{l}}$ is set to 1, otherwise it is 0:
	\begin{equation}
	\bar{\Lambda}^k_l =
	\begin{cases} 
	1 &   \Lambda^k_{l,\mathsf{RW}}  > {\lambda}\\
	0 & \text{otherwise}.
	\end{cases}
	\label{statistic}
	\end{equation}
	Hence, $\bar{\Lambda}^k_l$ indicates whether or not it is likely for the angle bin $l$ to contain a target.
	Hence state $s_k$ is then defined as the total number of angle bins where the targets could be located at time $k$:
	\begin{align}
	s_k &= \sum_{l=1}^L \bar{\Lambda}^k_{l}.
	 \label{state}
	\end{align}
	 Hence, the set of possible states can be written as $\mathcal{S}=\{0,\dots,M\}$.

	\subsection{The set of actions}
	The MIMO radar, i.e., the agent, starts by initially transmitting an orthonormal waveform at time $k=0$, by setting the beamforming matrix $\mathbf{ W }_k$ given in eq. \eqref{channel} to be equal to the identity matrix, i.e., $\mathbf{ W }_k=\mathbf{ I }$. We assume a discrete radar field of view divided into $L$ angle cells. Hence,  $\textbf{y}^k_{l}$ is the received signal of single snapshot at angle bin $l$.\\
	 The number of targets $s_{k+1}$ is calculated based on \eqref{state} which gives an indication about the status of the environment. Based on this observation, the agent takes a certain action. An action can be defined as two tasks. The first task is the process of selecting the candidate angle bins which most likely contain the targets based on the current environmental state. The second task is the optimization of the beamforming matrix, $\mathbf{ W }$, to focus the transmit power towards those angle bins. Hence, the cardinality of the set of actions $\mathcal{A}$ can be set as $M$, which is the maximum number of targets that the radar can detect. Consequently, an action can be defined as $a_k \in\mathcal{A}=\{\Theta_i|i\in\{1,2,\dots,M\}\}$. Specifically, the agent has to identify $\Theta_i=\{\hat{\theta}_1,\hdots,\hat{\theta}_i\}$ which  are the	$i$ angle bins that most likely contain targets. Here $\hat{\theta}$ is the estimated angle bin which contains a target, while $\mathcal{A}=\{\Theta_1,\Theta_2,\dots,\Theta_M\}$ is the set of all possible angle bins.  \\ In order to build the set $\Theta_i$, the highest $i$  values of  $ \Lambda^k_{l,\mathsf{RW}}$, defined in \eqref{lambda_k}, are chosen.
		In other words, we rank all the angle bins based on their $ \Lambda^k_{l,\mathsf{RW}}$ and the best $i$ angles are chosen. 
	With the completion of the last step, the agent is now ready to optimize the beamformers towards the chosen angles.
\\
	The weighting matrix $\mathbf{W}$ has to be optimized, in order to synthesize the corresponding beam pattern. Therefore, the transmitted power is concentrated towards those angle bins in $\Theta_i$, which may contain targets. \\ This is done by maximizing the minimum of the beam pattern $B(\hat\theta_j)=\mathbf{a}^T_T(\hat\theta_j)\textbf{R}_W\mathbf{a}^*_R(\theta_j)$, with $\textbf{R}_W=\textbf{WW}^H$, $\hat\theta_j\in\Theta_i$ and under the power constraint $\text{tr}(\textbf{R}_W)=P_T$. The resulting optimization problem is stated as follows:
	\begin{align}
	\label{optimprob}
	\text{max}_{\mathbf{W}}\text{min}_{j\in\mathcal{T}_i}\{\mathbf{a}^T_T(\hat\theta_j)\textbf{WW}^H \mathbf{a}^*_{R}(\hat\theta_j)\}\\
	\text{s.t. tr}(\textbf{WW}^H)=P_T,\notag
	\end{align}
	where $\mathcal{T}_i =\{1,\dots,i\}$. Details of the optimization problem and its solution are provided in Sect. \ref{Opt_Prob}.
	\subsection{Policy}
In order to determine which action the agent has to take at each time $k$, an action policy must be defined. This policy controls the size of $\Theta_i$ (i.e., $i$) which defines the action $a_k$. In our algorithm, $i$ is defined based on $\epsilon$ greedy policy, which is a simple policy balancing exploration and exploitation. In fact, the agent could follow the same actions which was tried in the past and proved to maximize the reward (i.e., exploitation). However, the agent is required also to acquire new knowledge through discovering new actions (i.e., exploration). In the  $\epsilon$ greedy policy, the variable $\epsilon$ refers to the probability of exploring new actions randomly. In more details, the optimal action $a_{\mathsf{opt}}\overset{\Delta}{=}\argmax_{a\in A}\mathbf{ Q }\left(s_{k+1},a\right)$ is taken with a probability of $1-\epsilon$, while another random action $a_{\mathrm{rnd}}$ (excluding $a_{\mathsf{opt}}$) is chosen with a probability of $\epsilon$ such that\\
	\begin{equation}
	{a}_{k+1} =
	\begin{cases} 
	a_{\mathsf{opt}} &  \mathrm{with}\hspace{0.05in}\mathrm{prob.} \hspace{0.05in} 1-\epsilon\\
	a_{\mathrm{rnd}} & \mathrm{with}\hspace{0.05in}\mathrm{prob.}\hspace{0.05in} \epsilon.
	\end{cases}
	\label{epsilon}
	\end{equation}
	This implies that, if we set $\epsilon=0$, the agent will not  explore anything and would always choose $a_{\mathsf{opt}}$. Whereas, if we set $\epsilon=1$, the action is selected randomly and the agent would 
	not exploit the information previously learned and saved in the state-action matrix $\mathbf{Q}$. In the following example, we further clarify how an action is defined.\\
	\begin{exmp}
		Suppose that the maximum number of targets that the radar can detect is set to $M=3$. Furthermore, $s_{k+1}=2$, which indicates that there are only two angle bins that most likely contain the target based on \eqref{state}. In addition, the $\epsilon$ greedy algorithm chose $a_{k+1}= a_{\mathsf{opt}}$ as per \eqref{epsilon} and the $\mathbf{ Q }$ matrix at time $k=17$ is given by Table 1.
		\begin{table}[!h]
			\centering
			\label{tab:my-table}
	\begin{tabular}{|l|l|l|l|l|}
		\hline
		\diagbox{States}{Actions} & $a_{0}$ & $a_{1}$ & $a_{2}$ & $a_{3}$ \\ \hline
		$s_{0}$  & 0.4508      & 0      & 0       & 0      \\ \hline
		$s_{1}$  & 1.4376    &2.2586    &1.5617    &2.4848      \\ \hline
		$s_{2}$  & 0.5118    &1.5951     &2.5540    &1.4495      \\ \hline
		$s_{3}$   & 0        & 0    &1.5345     &0        \\ \hline
	\end{tabular}
	\caption{$\mathbf{ Q }$ matrix  at time $k=17$}
\end{table}
Therefore, $a_{\mathsf{opt}}\overset{\Delta}{=}\argmax_{a\in A}\mathbf{ Q }\left(3,a\right)=a_2=\Theta_2$, as the maximum value in the third row is 2.5540 which corresponds to $\Theta_2=\{\hat{ \theta }_1,\hat{ \theta }_2\}$. To build $\Theta_2$, all the angle bins are ranked based on their $ \Lambda^k_{l,\mathsf{RW}} $ and the first two angles bins are selected. Those angles are accordingly used to optimize $\mathbf{ W }$ as in \eqref{optimprob}.
\begin{table}[!h]
	\centering
	\label{tab:my-table3}
	\begin{tabular}{|l|l|l|l|l|}
		\hline
		\diagbox{States}{Actions} & $a_{0}$ & $a_{1}$ & $a_{2}$ & $a_{3}$ \\ \hline
		$s_{0}$  & 0.4508       & 0      & 0       & 0      \\ \hline
		$s_{1}$  & 4.2332   & 2.0142   &4.3149   & 5.2843     \\ \hline
		$s_{2}$  &0.5118    &3.4302    &4.4225    &4.7489   \\ \hline
		$s_{3}$   & 1.8499        & 0    &1.5345    &2.3170        \\ \hline
	\end{tabular}
	\caption{$\mathbf{ Q }$ matrix  at time $k=50$}
\end{table}
	\end{exmp}
 In the first few time steps, $a_{\mathsf{opt}}$ might cause miss detection. This can be seen in the case of $k=17$, where if $s_{k+1}=3$, then the next action $a_{k+1}= a_{\mathsf{opt}}=a_2$ will be chosen with probability $1-\epsilon$. However, as the $\mathbf{ Q }$ matrix is updated every time step based on the new evaluated reward, the probability of miss detection decreases over time. Hence, in the final time step $k=50$, if $s_{k+1}=3$, then the next action $a_{k+1}= a_{\mathsf{opt}}=a_3$. 
	\subsection{The reward}
	The reward defines the goal of the RL problem, hence the radar agent's sole objective is to maximize the total cumulative reward function on the long run \cite{Sutton1998}. Consequently, it defines how the agent should behave, as the agent learns what are the good and bad actions. In our case, the goal is to detect all the targets even those masked within the disturbance. This is achieved through specific actions, i.e., optimizing the beampattern. Therefore, the reward is expressed in terms of the estimated  $	\hat{P}_{\mathsf{D}_l}^{k}$ as 
		\begin{equation}
		\hat{P}_{\mathsf{D}_l}^{k}=Q_1\left(\sqrt{\hat{\zeta}^k_l},\sqrt{\lambda}\right),
		\label{PD}
		\end{equation}
		\begin{equation}
		\hat{\zeta}^k_l=2|\hat{\alpha}^k_l|^{2}\frac{\norm{\mathbf{h}^k_l}^4}{(\mathbf{h}^k_l)^H\widehat{\mathbf{\Gamma}_l}\mathbf{h}^k_l},
		\label{zeta}
		\end{equation}
	
	where $Q_1\left(.,.\right)$ is the first order \textit{Marcum Q function} \cite{Nuttall},  $\widehat{\mathbf{\Gamma}}$ is defined as in \eqref{GammaHat}   
	and $\hat{\alpha}$ is an asymptotically Normal estimator of $\alpha$ which is $\sqrt N$ consistent, it is defined as
	\begin{equation}
	\hat{\alpha}^k_l=\frac{(\mathbf{h}^k_l)^H \mathbf{y}^k_l}{\norm{\mathbf{ h }^k_l}^2}.
	\label{alpha}
	\end{equation}
	The theoretical derivation of \eqref{PD} and \eqref{zeta} can be found in \cite{fortunati2019massive}. In fact, the reward is chosen to be function of \eqref{PD} as it provides accurate detection in an asymptotic regime when the number of spatial virtual antenna channels $N$ grows unbounded, i.e., $N\rightarrow \infty$. Furthermore, the robust Wald-type detection in eq. \eqref{statistic} satisfies the CFAR requirement even in a single-snapshot scenario.\\
	The reward function is composed of two parts: negative and positive reward.  The negative reward can be considered as penalty for the agent in case of false detections. Hence, the positive reward is the summation of $\hat{P}_{\mathsf{D}_l}^{k}$ for the angle cells defined in $i$, while the negative reward is the summation of  $\hat{P}_{\mathsf{D}_l}^{k}$ for the rest of the cells, which is likely not to contain any target. \\
	The reward for each time step $k$ will be defined as:
	\begin{align}
	r_{k+1} &= \sum^{s_k}_{l=1}\hat{P}_{\mathsf{D}_l}^{k}-\sum^{L-s_k}_{j=1}\hat{P}_{\mathsf{D}_j}^{k},
	\label{rewardeq}
	\end{align}
	where $\hat{P}_{\mathsf{D}_l}^{k}$ is the probability of detection as in \eqref{PD}
		calculated for target $l$ at $\theta_k^l$ after taking observing state $s_k$ and taking action $a_k$.\\ In the following algorithm, the steps of our MMIMO radar SARSA are summarized.
	\begin{algorithm}
				\caption{SARSA}
				\label{SARSA}
	\begin{algorithmic}
		\State Initialize $\mathbf{ Q }=\mathbf{0}_M$
		\State\text{Initialize } state $s_0=1$,  action $a_0=1$, $K=50$ and $\mathbf{W}_k=\mathbf{I}$
		\Repeat \text{ for each time step $k$:}
		\State \text{Take action $a_k$ by transmitting waveform \eqref{txSignal} using $\mathbf{W}_k$}
		\State Acquire the received signal $\mathbf{ y }^k_l$, $\forall$ $l =1,\dots, L$
		\State Calculate $s_{k+1}$ from \eqref{state}
		\State Evaluate the reward  $r_{k+1}$ as in \eqref{rewardeq}
		\State Choose action $a_{k+1}$ as \eqref{epsilon}, identify $\Theta_i$ and $\mathcal{ T }_i$
		\State \begin{align*}
		Q\left(s_k,a_k\right)\leftarrow &  Q\left(s_k,a_k\right)+\\& \alpha \left(r_{k+1}+ \gamma Q\left(s_{k+1},a_{k+1}\right) - Q\left(s_k,a_k\right)\right)\end{align*}
		\State $s_k\leftarrow s_{k+1}$;$a_k\leftarrow a_{k+1}$
			\If {$s_{k+1}$ $\neq$ 0}
		\State Solve for $\mathbf{W}_{k+1}$ in \eqref{optimprob} using algorithm \eqref{alg1}
			\Else
		\State $\mathbf{W}_k=\mathbf{I}$
		\EndIf 
		\Until  Observation time ends
	\end{algorithmic}
\end{algorithm}
	\section{Optimization Problem}

\label{Opt_Prob}
In this section, a solution for the beamforming optimization problem in  \eqref{optimprob} is discussed. As a matter of fact, semi-definite programming (SDP) relaxation is a widely used method to solve this problem \cite{Lipaper,asilomar,SDP}, however SDP complexity increases with the size of $\mathbf{W}$, hence using SDP for the MMIMO application previously described would not be realistic. Moreover, SDP involves a relaxation of the original problem and getting a feasible solution requires a heuristic randomization process. The high complexity of the solution described in \cite{ouini_BP,FR_BP,Lipaper} to the optimization problem is due to its two-steps structure: in the first step, $\textbf{R}_W$ is synthesized; then in the second step, the beamformer matrix $\mathbf{W}$ is generated from $\textbf{R}_W$.\\
To reduce the computational complexity, we propose another approach based on inner convex approximations (ICA) \cite{Nova}. This allows to find $\mathbf{W}$, in an iterative fashion. Our approach, as opposed to SDP, guarantees obtaining a Karush-Kuhn-Tucker (KKT) point of the original problem and avoids rank relaxation issues of the SDP approach. 
To this end, we write the optimization problem, \eqref{optimprob} as follows
\begin{align}\label{NewA}
& \max_{\mathbf{W}, \zeta}\zeta \\ & { \text { s. } t . } \quad {  \quad \zeta \geq 0 , \quad \operatorname { tr } \left(\mathbf {W}^H \mathbf {W}\right) = P _ { T } } \\ &  \quad \zeta - f_j(\mathbf {W}) \leq 0 \quad 
\forall j \in \mathcal { T } _ { i } .\label{Nonc} 
\end{align}
Here the function $f_j(\mathbf {W})$ is defined as $ f_j(\mathbf {W}) \triangleq { \mathbf { a }} _ { T } ^ { T } \left( \hat{ \theta } _ { j } \right) \mathbf {W} \mathbf {W}^H \mathbf { a } _ { R } ^ { * } \left( \hat { \theta } _ { j } \right)$. Problem \eqref{NewA} is non-convex and difficult to solve due to constraints in \eqref{Nonc} which are not convex. To overcome this difficulty, we propose to iteratively approximate the non-convex feasible set from inside with a convex feasible set by approximating the function $ f_j(\mathbf {W})$. The approximation of the function $f_j(\mathbf {W})$ writes
\begin{equation}
\tilde{f}_j(\mathbf {W}; \widetilde{\mathbf{W}}) = f_j(\widetilde{\mathbf{W}}) +\langle \nabla_{\mathbf {W}} f_j(\widetilde{\mathbf{W}}), \mathbf {W} - \widetilde{\mathbf{W}} \rangle,
\end{equation}
where $\nabla_{\mathbf {W}} f_j(\widetilde{\mathbf{W}})$ is the gradient of function $f_j(\mathbf {W})$ with respect to $\mathbf {W}$ computed at the fixed point $\widetilde{\mathbf{W}}$ and $\langle \mathbf{ A }, \mathbf{ B }\rangle = \rm{Re}\{\rm{tr}(\mathbf{ A }^H \mathbf{ B })\}$. Obviously, using this approach we get an inner convex approximation of the original non-convex feasible set. To see this, note that the function $f_j(\mathbf {W})$ is a quadratic convex function in $\mathbf {W}$ which results in its approximation being its under-estimator. Hence, we have the following relation
\begin{equation}
\zeta - f_j(\mathbf {W}) \leq \zeta - \tilde{f}_j(\mathbf {W};  \widetilde{\mathbf{W}}) \leq 0 \quad 
\forall j \in \mathcal { T } _ { i }
\end{equation}
This inner convex approximation approach, iteratively enhances the lower-bound on the convex function $ f_j(\mathbf {W})$ and eventually converges to a KKT point of problem \eqref{NewA}. Let $m$ be iteration index and $\widetilde{\mathbf {W}}^m$ be the beamforming matrix at iteration $m$.  The successive inner convex approximation approach is based on solving the following problem iteratively until convergence.
\begin{align}\label{NewAA}
& \max_{\mathbf{W}, \zeta}\zeta \\ & { \text { s. } t . } \quad {  \quad \zeta \geq 0 , \quad \operatorname { tr } \left(\mathbf {W}^H \mathbf {W}\right) = P _ { T } } \\ & \quad \zeta-  { \tilde{f}_j(\mathbf {W}; \widetilde{\mathbf {W}}^m) \leq 0 , \forall j \in \mathcal { T } _ { i }, } \label{Nonc1}
\end{align}
Problem \eqref{NewAA} is convex and the optimal solution can be found efficiently with an interior-point solver such as in \cite{cvx}. The algorithm for finding the KKT point of problem \eqref{NewA} is listed in algorithm \ref{alg1}.\\
\begin{algorithm}
	\caption{Iterative Inner Convex Approximation Algorithm}
	\label{alg1}
	\begin{algorithmic}
		\State Set $m = 0$ and initialize $\widetilde{\mathbf {W}}^0$ such that $\rm{tr}\left( (\widetilde{\mathbf {W}}^0)^H \widetilde{\mathbf {W}}^0\right) = \textit{P}_{T} $. 
		\State Repeat until convergence:
		\State Solve problem \eqref{NewAA} approximated around point $\widetilde{\mathbf {W}}^m$.
		\State Set $\widetilde{\mathbf {W}}^{m+1}$ as the optimal solution of problem \eqref{NewAA} 
		\State $m\leftarrow m+1$
	\end{algorithmic}
\end{algorithm}
\proposition 
Algorithm \ref{alg1} guarantees convergence to a KKT point of the non-convex problem \eqref{optimprob}.
\proof 
Please refer to the Appendix \ref{ApprendixB}.

	\section{Numerical analysis}
	In this section, the cognitive MIMO radar using the SARSA algorithm is simulated, where the agent is in a continuous learning mode of the surrounding environment, taking decisions while learning. The performance is averaged over $10^4$ Monte Carlo runs. Table \ref{RLParam} summarizes the values of the parameters for the SARSA algorithm.
	\begin{table}[H]\small
		\centering
		\begin{tabular}{p{2.8cm} p{0.7cm}|p{2.8cm} p{0.7cm}}
			\hline
			Parameter & Value & Parameter & Value\\
			\hline
			Learning rate $\alpha$  & 0.8 & Discount factor $\gamma$ & 0.8\\
			exploration rate $\epsilon_0$ & 0.5 & Time steps & 50 \\
			Number of states & 11& Initial state & 1 \\ Number of actions & 11	
		\end{tabular}
		\caption{Reinforcement learning parameters}
		\label{RLParam}
	\end{table}
	
	\subsection{Simulation Setup}
	In our simulations, we consider a uniform linear array (ULA) at the transmitter and receiver each with inter-element spacing of $d=\lambda/2$. The angle grid is divided into $L=20$ angle bins. The angular locations would be represented in terms of the spatial frequency $\nu$, which is defined as
		\begin{equation}
	\nu\overset{\Delta}{=} \frac{d f_c}{c} \sin\left(\theta\right)
	\end{equation}
	where $f_c$ is the carrier frequency and $c$ is the speed of light. Hence, the steering vector for the transmit or receive can be redefined in terms of $\nu$
	\begin{equation}
	\mathbf{a}_R(\theta) = [1,e^{j 2\pi \nu},\dots,e^{j 2\pi (N-1)\nu}]^T,
	\end{equation} 
		where $N$ is the number of transmit or receive antennas.
	Furthermore, the angle grid can be expressed as a spatial frequency grid where $\boldsymbol{\nu}=\left[-0.5:0.45\right]$.
	 We further assume the existence of four targets at spatial frequency locations $\nu=\{-0.2,0,0.2,0.3\} \subset \boldsymbol{\nu}$, with $\mathrm{SNR}=[-5\mathrm{dB},-8\mathrm{dB},-10\mathrm{dB},-9\mathrm{dB}] $ respectively.

	
	\subsection{Disturbance Model}
\begin{figure}
		\centering
			\includegraphics{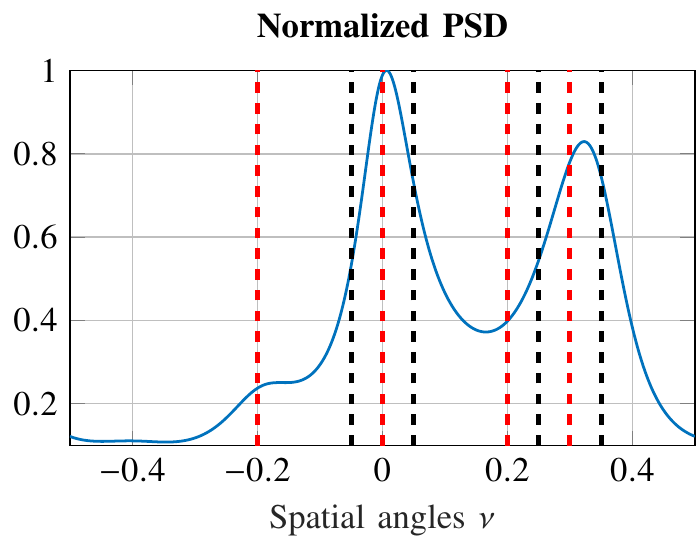}
		\caption{Disturbance PSD along with targets angles locations (red dashed lines for study case 1 and black ones for study case 2)}
		\label{PSD clutter}
	\end{figure}

	The disturbance model is chosen to mask the target angles, where the disturbance power is spread all over the spatial frequency range. Hence, the potential of our RL cognitive radar algorithm can be analyzed in such harsh environment.
	The disturbance $\mathbf{c}^k_l$ is generated according to the model of circular SOS $\mathrm{AR}\left(p\right)$ \cite{fortunati2019massive} as
	\begin{equation}
	c_n=\sum_{i=1}^{p} \rho_i c_{n-i}+w_n, \hspace{0.1in} n \in \left(-\infty,\infty\right),
	\end{equation}
	where $p=6$, driven by identically independent (i.i.d.), $t$-distributed innovations $w_n$ whose pdf $p_w$ is defined as \cite{Stefano_AR,fortunati2019massive} :
	\begin{equation}
	p_w\left(w_n\right)=\frac{\mu}{\sigma^{2}_w}\left(\frac{\mu}{\xi}\right)^{\mu}\left(\frac{\mu}{\xi}+\frac{\left | w_n\right|^{2}}{\sigma^{2}_w}\right)^{-\left(\mu+1\right)}.
	\end{equation}
	$\mu \in \left(1,\infty\right)$ is the shape parameter controlling the non-Gaussianity of $w_n$. Specifically, if $\mu \rightarrow 1$, then $p_w$ is a heavy tailed pdf with highly non-Gaussian behavior. However, the pdf becomes Gaussian if $\mu \rightarrow \infty$. The scale parameter is defined by $\xi=\mu / \left(\sigma^{2}_w\left(\mu-1\right)\right)$. We set in our simulations $\mu=2$ and $\sigma^2_w=1$.
	 Hence, the normalized power spectral density (PSD) of the disturbance is given by \cite{fortunati2019massive}
	 	\begin{equation}
	 	S(\nu)\overset{\Delta}{=}\sigma^{2}_w \left | 1-\sum_{n=1}^{p}\rho_n e^{-j2\pi\nu} \right | ^{-2}, \hspace{0.2in} p=6.
	 	\end{equation}
The coefficient vector $\rho$ is defined as
\begin{align}
\rho=&[0.5e^{-j2\pi0.4},0.6e^{-j2\pi0.2},0.7e^{-j2\pi0},0.4e^{-j2\pi0.1},\\&0.5e^{-j2\pi0.3},0.6e^{-j2\pi0.35}]^T \nonumber.
\end{align} 
The disturbance PSD is shown in Fig.\ref{PSD clutter}, where the target angles are marked in red dashed lines. Note that the disturbance PSD has multiple peaks.
	\subsection{Study Case 1 : Stationary Environment}
	To exploit the benefits of RL, we compared the the proposed RL-based beamformer against  two types of beamforming, \begin{itemize}
			\item omnidirectional equal power allocation with no RL. Here, the antennas emit orthogonal waveforms and the power is divided equally across all antennas
			\item adaptive beamforming represented by Alg (2) with no RL.
		\end{itemize}
	 It is assumed that the total power $P_T =1$ and for fair comparison the same detector $\Lambda^k_{\mathsf{RW}_l}$ is used in all cases according to \eqref{statistic} in each time step. In this set of simulations, the environment is assumed to be temporally stationary.
	The difference in behavior of the MMIMO radar with / without RL is analyzed. The results were averaged over $10^4$ Monte Carlo runs \footnote{For reproducible research, the source code is available for download at https://gitlab.com/aya.ahmed/rl-based-approach-for-mimo-radar-detection.git}.
	\subsubsection{Scenario 1}
\begin{figure*}[h!]
		\begin{subfigure}[b]{0.5\textwidth}
				\centering
				\includegraphics{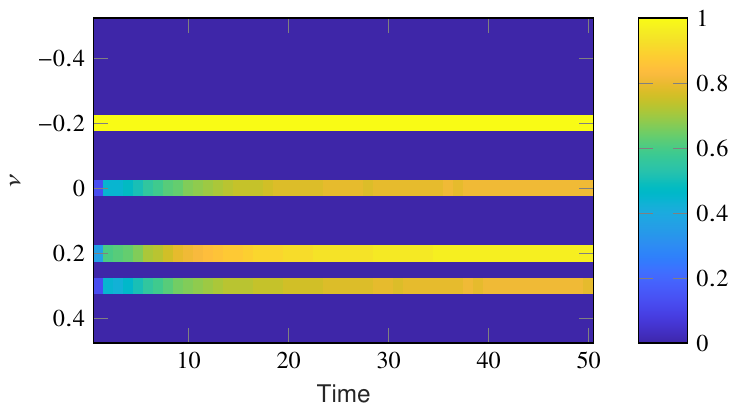}
			\caption{RL based beamforming}
				\label{RLdetection}
		\end{subfigure}
		~ 
		\quad
		\begin{subfigure}[b]{0.5\textwidth}
			\centering
				\includegraphics{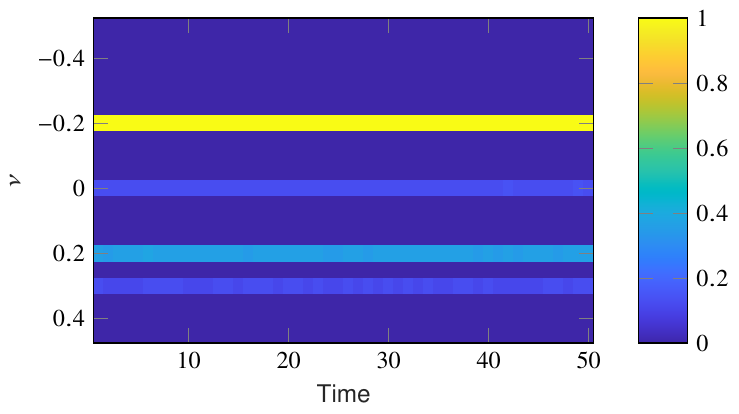}
			\caption{Omni-directional with equal power allocation }
				\label{omnidetection}
		\end{subfigure}
	\caption{Detection performance of proposed RL beamforming vs. omnidirectional with equal power allocation under $P_{\mathsf{FA}}=10^{-4}$ and $N=N_T N_R=10^4$. }
	\label{detectionComparison}
	\end{figure*}

\begin{figure}[!h]
			\centering
			\includegraphics{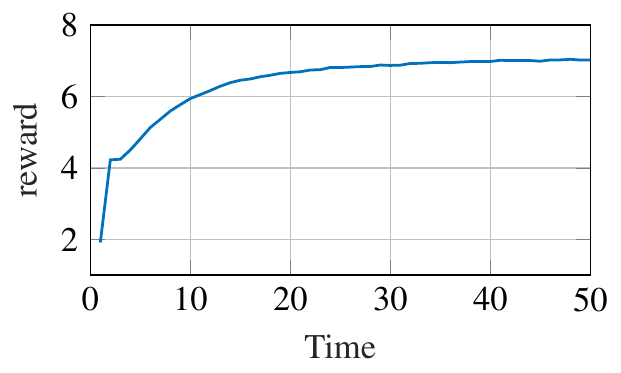}
			\caption{Reward calculated as in \eqref{rewardeq} for RL-based beamformer with $P_{\mathsf{FA}}=10^{-4}$ and $N=N_T N_R=10^4$. }
			\label{reward1}
	\end{figure}

\begin{figure}[!h]
	\centering
		\includegraphics{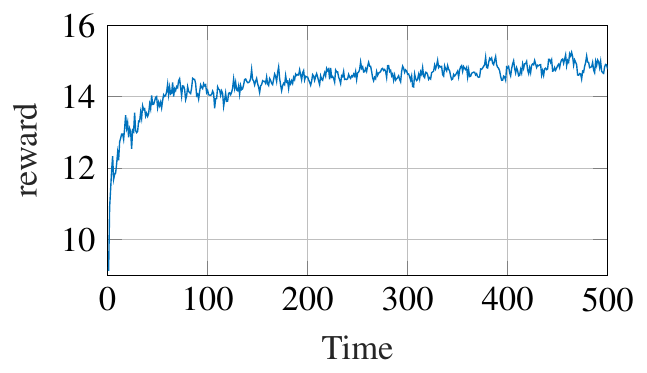}
	\caption{Reward behavior in static environment for 9 targets. }
	\label{reward2}
\end{figure}

	In this scenario, the performance of the algorithm is analyzed for a MMIMO regime where $N=N_T N_R=10^4$ and $P_{\mathsf{FA}}=10^{-4}$. Fig. \ref{detectionComparison} depicts the difference between our proposed algorithm and omnidirectional MIMO. In order to obtain those figures, we calculated the threshold in \eqref{statistic} within each time step, then the average is taken across all Monte Carlo runs. Fig. \ref{RLdetection} demonstrates better detection performance for all targets even the ones with low $\mathrm{SNR}$. It can be shown that the algorithm learns across time: in the first ten time steps, the agent is learning the disturbance, enhancing its experience as the time passes. Conversely, in the omnidirectional approach in Fig. \ref{omnidetection} the targets with lower $\mathrm{SNR}$ are mostly masked under the disturbance peaks, as in the case of $\nu=0$ and  $\nu=0.3$.
	\\
To measure the convergence of our algorithm, we report the immediate reward function as in \eqref{rewardeq} in Fig. \ref{reward1}. In fact, as depicted from the figure, the reward converges after 20 time steps. This result is consistent with Fig. \ref{RLdetection}, where the agent performance becomes very good after $k=20$. However, it is worth mentioning, that the convergence of the reward depends heavily on the environment and the number of targets in the scene with respect to the size of the $Q$ matrix. To verify this claim, we further increased the number of targets to 9 while keeping the same size of the $Q$ matrix. It can be shown from Fig. \ref{reward2}, that, in this case, the convergence is reached after 150 time steps $k=150$. This is due to the fact that the $Q$ matrix becomes less sparse as the number of targets increase w.r.t the matrix size.
	\subsubsection{Scenario 2}
\begin{figure*}[t!]
		\begin{subfigure}[b]{0.5\textwidth}
			\includegraphics{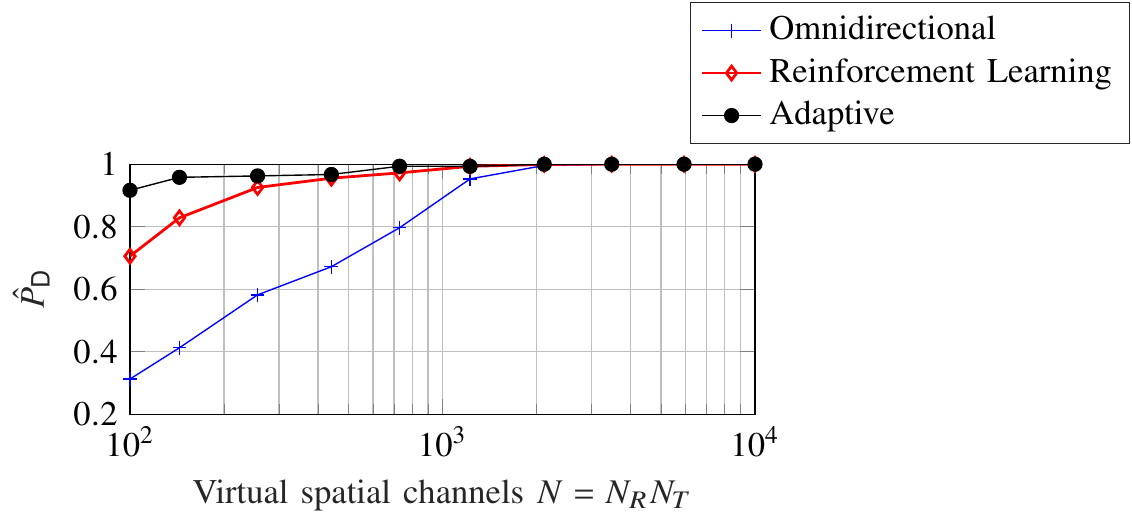}
			\caption{$\nu = -0.2$}
				\label{-0.2PDAnt}
		\end{subfigure}
		~ 
		\quad
		\begin{subfigure}[b]{0.5\textwidth}
			\centering
			\includegraphics{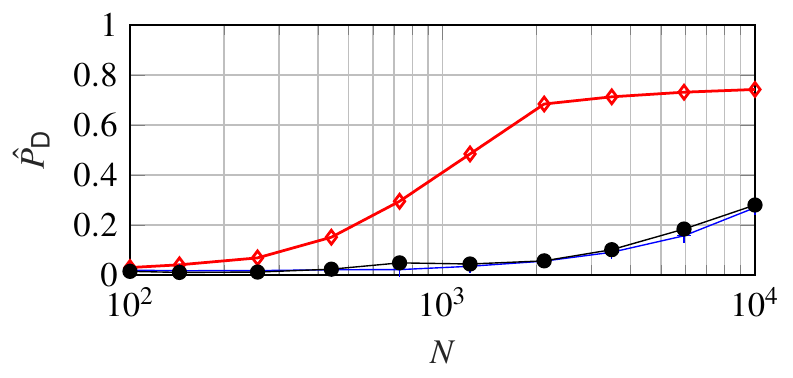}
			\caption{$\nu = 0$}
				\label{0PDAnt}
		\end{subfigure}
		~ 
		\quad
		\begin{subfigure}[b]{0.5\textwidth}
			\includegraphics{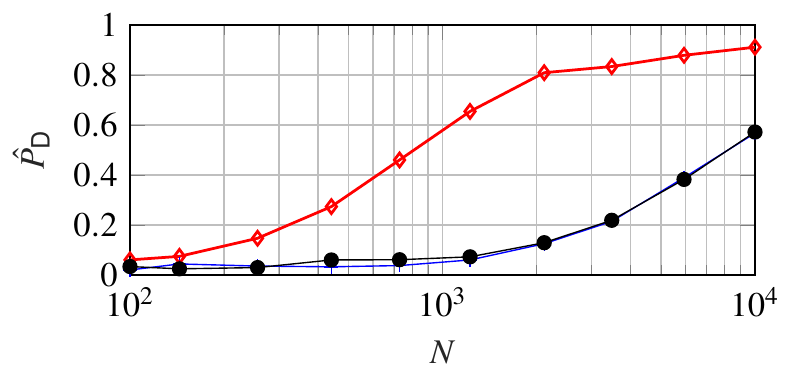}
			\caption{$\nu = 0.2$}
				\label{0.2PDAnt}
		\end{subfigure}
		~ 
		\quad
		\begin{subfigure}[b]{0.5\textwidth}
			\centering
			\includegraphics{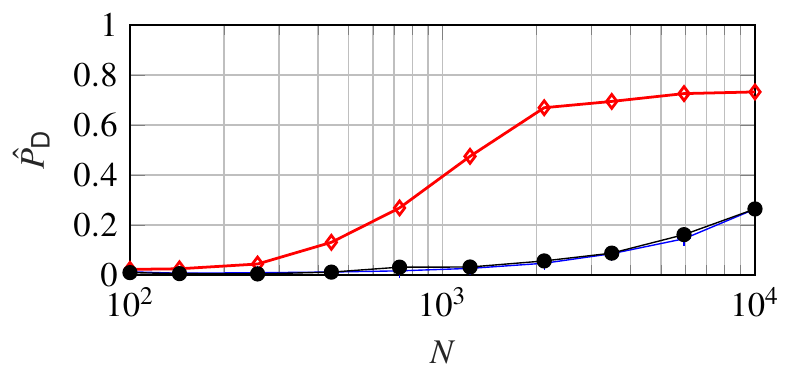}
			\caption{$\nu = 0.3$}
				\label{0.3PDAnt}
		\end{subfigure}
	\caption{$\hat{P}_\mathsf{D}$ using RL and alternative approaches of existing targets across different virtual antenna array size with $P_{\mathsf{FA}}=10^{-4}$ a) $\nu=-0.2$ with $\mathrm{SNR}=-5$ dB b) $\nu=0$ with $\mathrm{SNR}=-8$ dB  c) $\nu=0.2$ with $\mathrm{SNR}=-10$ dB d) $\nu=0.3$ with $\mathrm{SNR}=-9$ dB.}
		\label{PDvsN}
	\end{figure*}

	In this scenario we simulated the $\hat{P}_{\mathsf{D}}$  estimated from the closed form expression in \eqref{PD}, averaged over time for each target, as a function of the spatial virtual antenna channels $N$. Here, the number of transmit and receive antennas are $N_T$ $=$ $N_R$ $=$ $[10,12,16,21,27,35,46,59,77,100 ]$.\\The results in Fig. \ref{PDvsN} show that as $N$ increases, the $\hat{P}_\mathsf{D}$ increases for all the targets. However, the suggested algorithm provides better performance than the omni-directional and adaptive case for all targets except $\nu=-0.2$ as in Fig.~\ref{-0.2PDAnt}. In this case, the adaptive algorithm shows the best performance. This is due to the high SNR of this target, in addition it lies within low disturbance. The low $\hat{P}_\mathsf{D}$ as $N\rightarrow 10^3$ shown in Fig. \ref{0PDAnt}, \ref{0.2PDAnt} and \ref{0.3PDAnt}, is due to the harsh operating conditions, since $P_T=1$ and the nominal $P_{\mathsf{FA}}=10^{-4}$. Furthermore, the corresponding targets are located within heavy disturbance.
\subsubsection{Scenario 3}	
\begin{figure*}[t!]
		\begin{subfigure}[b]{0.5\textwidth}
			\includegraphics{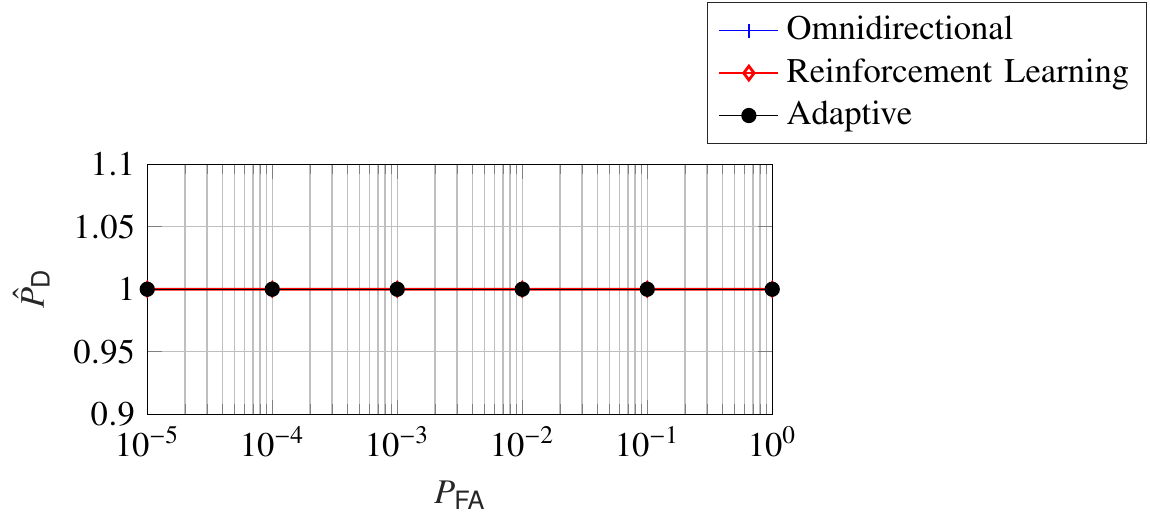}
			\caption{$\nu = -0.2$}
				\label{-0.2PDPF}
		\end{subfigure}
		~ 
		\quad
		\begin{subfigure}[b]{0.5\textwidth}
			\centering
			\includegraphics{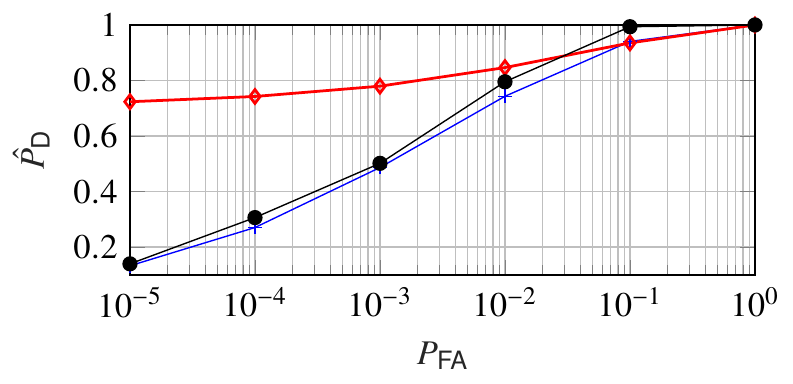}
			\caption{$\nu = 0$}
				\label{0PDPF}
		\end{subfigure}
		~ 
		\quad
		\begin{subfigure}[b]{0.5\textwidth}
			\includegraphics{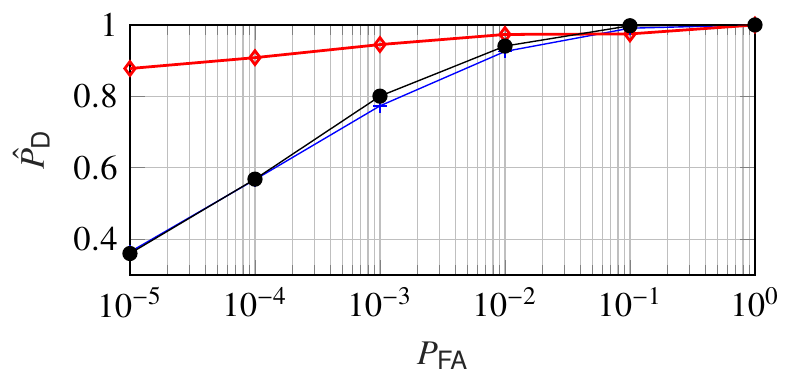}
			\caption{$\nu = 0.2$}
				\label{0.2PDPF}
		\end{subfigure}
		~ 
		\quad
		\begin{subfigure}[b]{0.5\textwidth}
			\centering
			\includegraphics{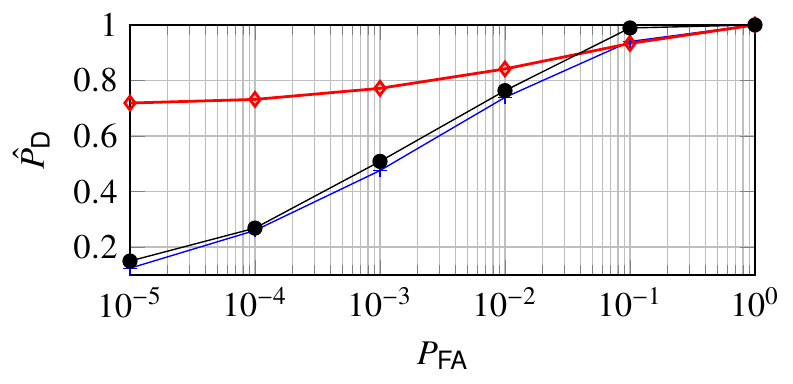}
			\caption{$\nu = 0.3$}
				\label{0.3PDPF}
		\end{subfigure}
		\caption{$\hat{P}_\mathsf{D}$ using RL and alternative approaches of existing targets across different $P_{\mathsf{FA}}$  with $N=$ $10^4$ a) $\nu=-0.2$ with $\mathrm{SNR}=-5$ dB b) $\nu=0$ with $\mathrm{SNR}=-8$ dB  c) $\nu=0.2$ with $\mathrm{SNR}=-10$ dB d) $\nu=0.3$ with $\mathrm{SNR}=-9$ dB. }
		\label{ROC}
	\end{figure*}

In this scenario, the receiver operating characteristics curve (ROC) is simulated across a range of $P_{\mathsf{FA}}=[10^{-5},10^{-4},10^{-3},10^{-2},1]$ with $N=10^4$
As shown in Fig. \ref{ROC}, mainly the potential of the RL-based cognitive MMIMO radar is shown in low $P_{\mathsf{FA}}$ regimes. As a matter of fact most practical radar applications has to maintain preassigned low $P_{\mathsf{FA}}$ values. Hence, we conclude that our proposed algorithm is more suitable for those practical systems in general.  Meanwhile, it is notably visible from Fig. \ref{-0.2PDPF}, that the $\hat{P}_\mathsf{D}$ is 1 across all $P_{\mathsf{FA}}$. This is due to the fact that this target has relatively high $\mathrm{SNR}$ and located within relatively low disturbance PSD. This means that omnidirectional and adaptive systems can perform well in those conditions. Meanwhile, the $\hat{P}_\mathsf{D}$ for targets with low $\mathrm{SNR}$ is much higher for the proposed algorithm compared to the omni directional and adaptive solutions, i.e., $\nu=0$ in Fig. \ref{0PDPF},  $\nu=0.2$ in Fig. \ref{0.2PDPF}. For both algorithms, $\hat{P}_\mathsf{D}$ approaches 1 for as the $P_{\mathsf{FA}}\rightarrow 1$ .
%
	\subsection{Dynamic Environment}
	In these simulations, the environment changes and the performance of our algorithm is analyzed such that the radar agent capability to adapt to those changes is tested. The number of total time steps is 100 and the results are averaged over 1000 Monte Carlo runs.
	\subsubsection{Scenario 4 : Changing Spatial frequencies}
\begin{figure*}[h!]
		\begin{subfigure}[b]{0.5\textwidth}
			\centering
				\includegraphics{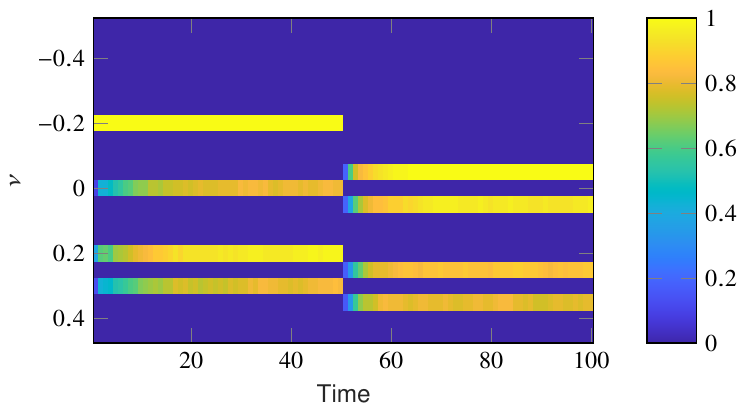}
			\caption{RL based beamforming}
			\label{RLangles}
		\end{subfigure}
					~ 
	\quad
		\begin{subfigure}[b]{0.5\textwidth}
			\centering
				\includegraphics{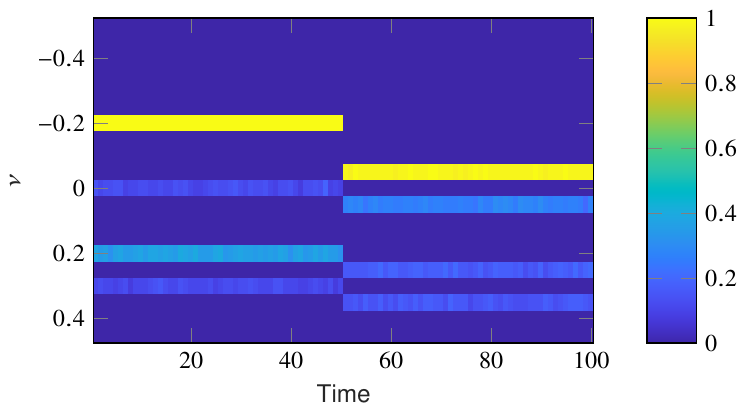}
			\caption{Omnidirectional with equal power allocation}
				\label{omniangles}
		\end{subfigure}
	\caption{Detection performance of RL based beamforming vs omnidirectional with equal power allocation for dynamic environment: changing angles at $T=50$.}
	\label{angleschange}
	\end{figure*}

\begin{figure}
	\centering
		\includegraphics{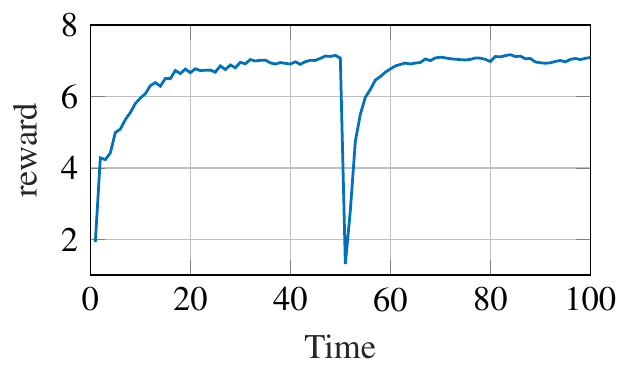}
	\caption{RL reward in dynamic environments: changing angles at $T=50$.}
	\label{reward_angle_change}
\end{figure}

	In this scenario, the targets' spatial frequencies are changed after 50 time steps. In this case $\nu$ is changed from $[-0.2,0,0.2,0.3]$ to $[-0.05,0.05,0.25,0.35]$ where the new spatial frequencies are depicted in dashed blue in Fig.\ref{PSD clutter}. In this case, we aim at  simulating a dynamic environment as shown in Fig. \ref{angleschange}, while their respective $\mathrm{SNR}$ remains the same, where $N=10^4$ and $P_{\mathsf{FA}}=10^{-4}$. On one hand, Fig. \ref{omniangles} shows the performance of the omnidirectional case, where it can detect only  targets whose new spatial frequencies are located where the disturbance PSD is low, i.e., $\nu =0.1$. On the other hand, the RL based beamforming algorithm can detect all the targets, even those lying close to the disturbance PSD peaks.\\
	Similar to scenario 1, Fig. \ref{reward_angle_change} shows the reward behavior across time as calculated in \eqref{rewardeq} and averaged over the Monte Carlo runs. The reward shows convergence after $T=20$ time steps. Then when the environment is changed by changing the angles, the agent sensed that change through exploration. Hence, the drop in the reward is seen after $T=50$ time steps, where the agent starts re-learning the changes, then after 10 time steps, the reward converges again.
\subsubsection{Scenario 5: Fading Targets}
\begin{figure*}[!h]
      	\begin{subfigure}[b]{0.5\textwidth}
      		\centering
      			\includegraphics{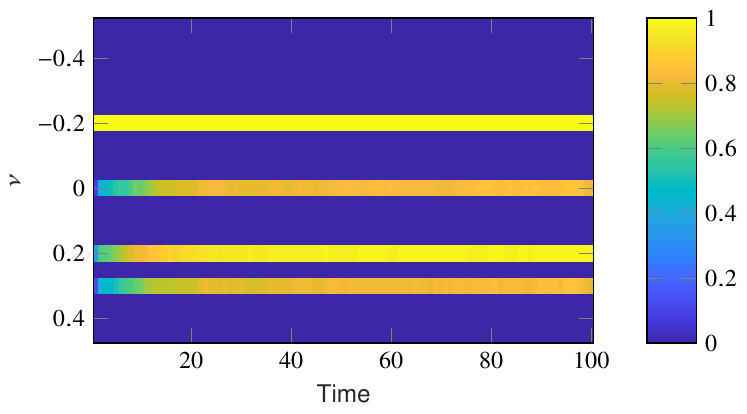}
      		\caption{RL based beamforming}
      		\label{RL_fading}
      	\end{subfigure}
      	~ 
      \quad
		\begin{subfigure}[b]{0.5\textwidth}
			\centering
				\includegraphics{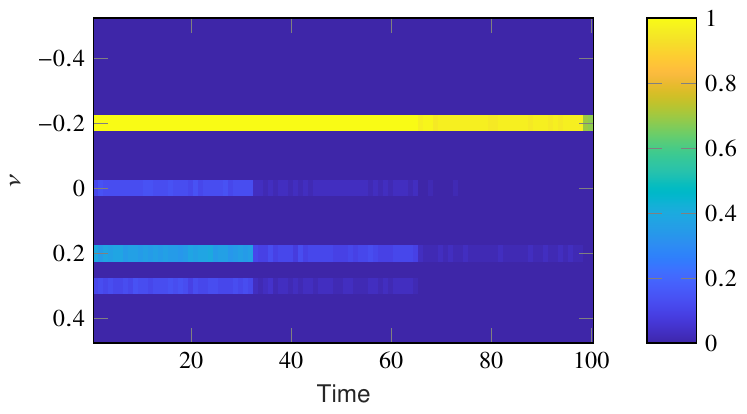}
			\caption{Omnidirectional with equal power allocation}
				\label{omni_fading}
		\end{subfigure}
	\caption{Detection performance of RL based beamforming vs omnidirectional with equal power allocation for dynamic environment i.e., target's $\mathrm{SNR}$ decreases by 20 \% every 30 time steps.}
		\label{detection_fading}
	\end{figure*}

\begin{figure}
		\centering
			\includegraphics{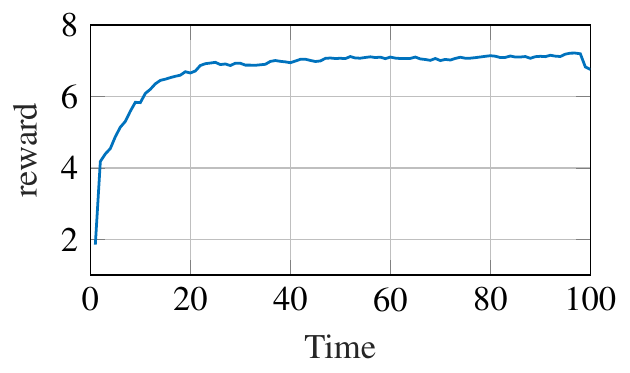}
		\caption{RL reward in dynamic environments: fading targets.}
			\label{reward_fading}
	\end{figure}

Here we simulate a different change in the environment, where we assume that the targets are fading and their $\mathrm{SNR}$ is decreasing. In Fig. \ref{detection_fading}, the targets' $\mathrm{SNRs}$ are assumed to decrease by 20\% every 30 time steps. Hence, by $T=90$, all targets' $\mathrm{SNRs}$ would have decreased by 60\%. In Fig. \ref{omni_fading}, the omnidirectional approach could not detect most of the targets after the first 30 time steps. Furthermore, the first target located at $\nu=-0.2$, which proved very good performance in the previous simulations due to its good $\mathrm{SNR}$ starts fading at $T=90$. This proves that the omnidirectional approach fails in the fading scenarios, since the radar here does not learn anything from the environment unlike in the RL case. Hence, it can not adapt to such changes. However, our proposed RL based beamforming algorithm obviously proves to have a reliable performance across the entire time steps. It can be concluded that RL cognitive MMIMO radar can adapt very well to all the environmental changes with very good performance. The corresponding reward behavior is shown in Fig. \ref{reward_fading}, where the algorithm can adapt to those changes in the $\mathrm{SNR}$ without any convergence issues.

	{\subsubsection{Scenario 6: Changing number of targets}
 In this scenario, we simulated total time of $K=60$ pulses, with $N=10^4$, $P_T=1$ and $P_\mathsf{FA}=10^{-4}$. The targets angles and SNR change over five time intervals $[1,\dots ,10, 11,\dots ,20, 21, \dots ,35, 36, \dots ,50, 51, \dots ,60]$ as below
 \begin{enumerate}
 \item $k=1\rightarrow10$, two targets at angles $\nu=\{-0.2,0\}$ with respective SNR of -5 and -9 dB.
 \item  $k=11\rightarrow20$,  no targets at all
 \item $k=21\rightarrow35$, five targets at angles $\nu=\{-0.2,0,0.15,0.25,0.3\}$ with respective SNR of -6, -8, -10, -11 and -12 dB
 \item $k=36\rightarrow50$, two targets at angles $\nu=\{0,0.2\}$ with respective SNR -9 and -8 dB.
 \item  $k=51\rightarrow60$ there exist 4 targets at angles $\nu=\{-0.05,0.05	,0.15,0.25,0.35\}$ with respective SNR -8,-7,-10 and -13 dB.
 \end{enumerate}
 As depicted in Fig. \ref{fig:changingtargets}, the RL beamformer is able to track all the targets, while the orthogonal beamformer failed to track most of the targets whose SNR is low or hidden within clutter. In more details, in the time interval (1),  both approaches can detect $\nu=0.2$ as it has high SNR unlike the case when $\nu=0$, where the RL approach shows better detection. \\ In time interval (2) both algorithms don't detect any targets due to the accuracy of the detection statistic used $\Lambda^k_{\mathsf{RW}_l}$ as $N$ grows asymptotically. In time interval (3), the omnidirectional approach faces difficulties detecting angles $\nu=\{0,0.3\}$ as they lie within high clutter PSD. In addition, it also fails to detect angle $\nu=0.25$ although it lies within low clutter PSD, due to the fact that this angle has very low SNR $=-11$ dB. In time interval (4) and (5), similar behavior is observed, where the omnidirectional approach fails to detect targets having low SNR or lying within high clutter PSD. However, our RL beamformer can even detect targets suffering from both problems at the same time as the case in time interval (5) with angle $\nu=0.35$.\\
  \begin{figure*}
	\begin{subfigure}[b]{0.5\textwidth}
		\centering
			\includegraphics{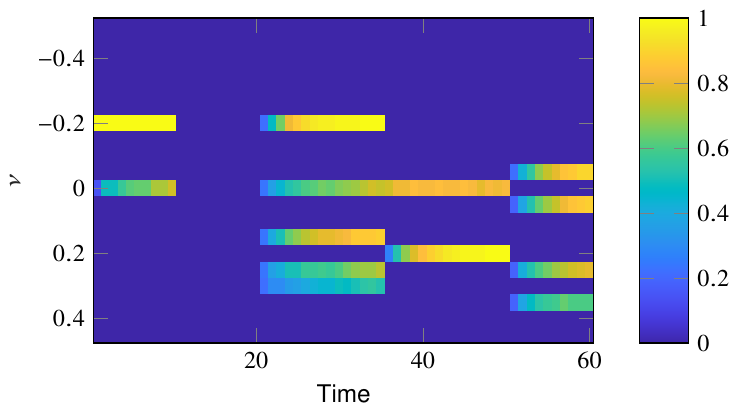}
		\caption{RL based beamforming}
		\label{fig:changingtargetsRL}
	\end{subfigure}
	~ 
	\quad
	\begin{subfigure}[b]{0.5\textwidth}
		\centering
		\includegraphics{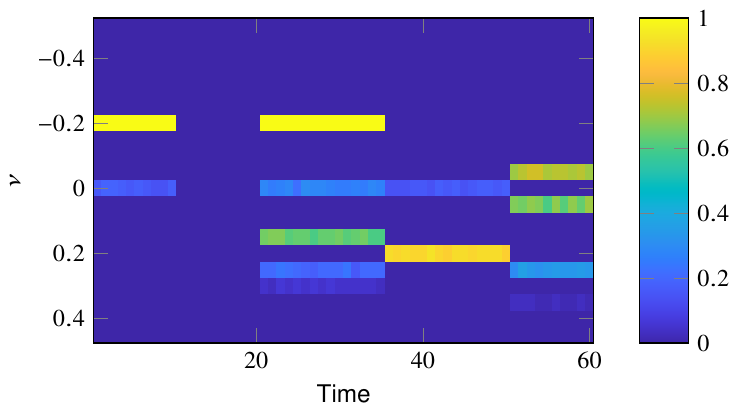}
		\caption{Omnidirectional with equal power allocation}
		\label{fig:changingtargetsOrth}
	\end{subfigure}
	\caption{Detection performance of RL based beamforming vs omnidirectional with equal power allocation for dynamic environment i.e., target's $\mathrm{SNR}$ and number changes over five time intervals.}
		\label{fig:changingtargets}
\end{figure*}
\subsubsection{Computational complexity}
In order to evaluate the complexity of the optimization algorithm proposed in Algorithm (2), the computational time/complexity is measured using the Matlab tic-toc function. This is done for both our algorithm and the conventional SDP solution presented in \cite{SDP} using the same simulation conditions. Fig.~\ref{fig:optimization} shows both solutions achieve similar performance for low $N_T$ values. However as the number of transmit antennas $N_T$ increases asymptotically, our proposed algorithm shows much less time complexity compared to the SDP solution. Hence, it can be deduced that Algorithm (2) is highly efficient for massive MIMO regimes. 
\begin{figure}
	\centering
		\includegraphics{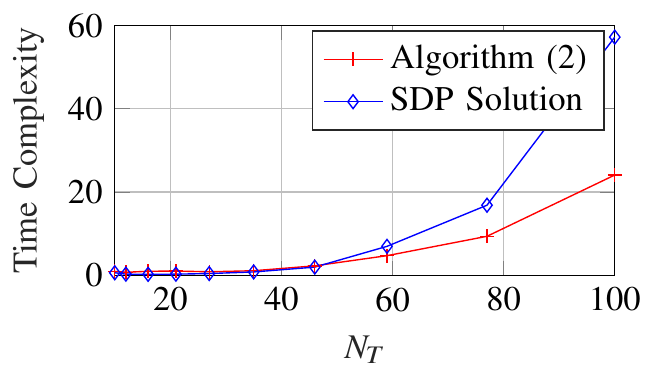}
	\caption{Performance evaluation of proposed optimization algorithm (2) and SDP solution as in \cite{SDP}  }
	\label{fig:optimization}
\end{figure}

	\section{Conclusion}
	In this paper, we studied the problem of multi-target detection for a MMIMO CR in the presence of unknown disturbance. We proposed a novel RL based beamforming algorithm, which could detect the targets with very low $\mathrm{SNR}$ even if the environment is dynamic. Specifically, the CR acted as an agent sensing the unknown environment (i.e., targets and disturbance) through illuminating it by transmitting a set of waveforms. Afterwards, a reward function is calculated from the reflected echoes. This reward has been defined as the closed form asymptotic expression of the $\hat{P}_\mathsf{D}$ as the number of virtual spatial antenna channels $N$ go to infinity that is the Massive MIMO regime \cite{fortunati2019massive}. The agent's goal is to maximize the reward through a course of actions without any a priori knowledge about the disturbance distribution, nor the targets number. In our case, those actions were tailoring the beampattern by optimizing the beamformer according to the acquired knowledge. \\
	Furthermore, we presented a novel approach for beamforming optimization, which is scalable as the size of $N$ increases and does not increase the complexity. Our numerical results showed a really good $ \hat{P}_\mathsf{D}$ performance for our algorithm as $N\rightarrow 10^4$ compared to the adaptive and omnidirectional approaches. In addition, the ROC confirmed the advantages of adopting a RL-based approach when the targets are embedded in spatially correlated heavy-tailed disturbance. The probability of detecting the low- $\mathrm{SNR}$ targets improves significantly. Moreover, a dynamic environment has been simulated by changing target angles, number and simulating target fading. In both cases, the proposed RL-based beamformer is able to adapt to the fast changing environment, without any a priori knowledge and to provide better performance than the classical (omnidirectional) beamformer.\\ This work focused on the multiple targets detection problem. Future work will investigate how to refine the direction of arrival (DOA) estimate of the detected targets within the RL framework applied here, in the presence of disturbance with unknown distribution.
	\appendix
	\subsection{Appendix A}
	In general, the disturbance statistics $\mathbf{\Gamma}$ is unknown. Hence according to\cite[Remark 1]{fortunati2019massive}, an estimate of the entries of $\mathbf{\Gamma}$ are given by
	\begin{equation}
	\left[\widehat{\Gamma}_l \right]_{i,j}=\left\{\begin{matrix}
	\hat{c}_i \hat{c}^*_{j} & j-i\leq l \\ 
	\hat{c}^*_i \hat{c}_{j} & i-j\leq l  \\ 
	0 & \left | i-j \right | > l
	\end{matrix}\right.
	\label{GammaHat}
	\end{equation}
where $l$ is the truncation lag \cite{Bdefin} and $\hat{c}=y_n-\hat{\alpha}h_n$.
	Generally speaking, if Assumption 1 holds, then the asymptotic distribution under $H_0$ and $H_1$ of the Wald statistic is
	
	\begin{align}
		\Lambda^k_{l}  & \left( \mathbf { y } _ { l , g } ^ { k } | H _ { 0 } \right) \underset { N _ { T } N _ { R } \rightarrow \infty } { \stackrel { d } { \sim } } \chi ^ { 2 } _ { 2 }\left(0\right),&\\
		\Lambda^k_{l}  & \left( \mathbf { y } _ { l , g } ^ { k } | H _ { 1 } \right) \underset { N _ { T } N _ { R } \rightarrow \infty } { \stackrel { d } { \sim } } \chi ^ { 2 } _ { 2 }\left(\zeta\right),
	\label{ARW}
	\end{align}
	where 
	\begin{equation}
	\zeta=2|\alpha|^{2}\frac{\norm{\mathbf{h}}^4}{\mathbf{h}^H\mathbf{\Gamma}\mathbf{h}}.
	\end{equation}
	Furthermore, under asymptotic detection performance of \eqref{ARW}, a closed form expression for $P_\mathsf{D}$ can be formulated as
	\begin{equation}
	P_\mathsf{D}\left(\lambda\right) \rightarrow_{N\rightarrow\infty}Q_1\left(\sqrt{\zeta},\sqrt{\lambda}\right),
	\label{PDAsymptotic}
	\end{equation}
	where $Q_1\left(\cdot,\cdot\right)$ is first order \textit{Marcum Q function} \cite{Nuttall}.
		\subsection{Appendix B}
		\label{ApprendixB}
		Proof of Proposition 1: 
		Let us write the non-convex constraint \eqref{Nonc}in problem \eqref{NewA}
		as 
		\begin{equation} \label{nonC}
		g_j(\zeta, \mathbf {W}) \leq 0 \quad 
		\forall j \in \mathcal { T } _ { i }
		\end{equation}
		where $g_j(\zeta, \mathbf {W}) \triangleq \zeta - f_j(\mathbf {W})$. In the approximate problem, we replace the constraint as written in \eqref{nonC} with the constraint: 
		\begin{equation} 
		\tilde{g}_j(\zeta, \mathbf {W}; \widetilde{\mathbf{W}}^m ) \leq 0 \quad 
		\forall j \in \mathcal { T } _ { i }
		\end{equation}
		such that $\tilde{g}_j(\zeta, \mathbf {W}; \widetilde{\mathbf{W}}^m) \triangleq \zeta - {\tilde{f}_j(\mathbf {W}; \widetilde{\mathbf {W}}^m) \leq 0 , \forall j \in \mathcal { T } _ { i }}$. 
		The function $\tilde{g}_j(\zeta, \mathbf {W}; \widetilde{\mathbf{W}}^m)$ satisfy the following properties: 
		\begin{itemize}\label{Prop}
			\item[1] The function  $\tilde{g}_j(\zeta, \mathbf {W}; \widetilde{\mathbf{W}}^m)$ is differentiable for all the values of $\left\lbrace \zeta, \widetilde{\mathbf{W}}^m\right\rbrace \in \mathcal{ F}$
			where, $\mathcal{ F} = \left\lbrace \zeta, {\mathbf{W}}| \quad \zeta \in \mathbb{R}_+, \quad \operatorname { tr } \left(\mathbf {W}^H \mathbf {W}\right) = P _ { T } \right\rbrace $
			\vspace{3mm}
			\item[2] $g_j(\zeta, \mathbf {W}) \leq  \tilde{g}_j(\zeta, \mathbf {W}; \widetilde{\mathbf{W}}^m) \quad \forall \hspace{2mm} \widetilde{\mathbf{W}}^m \in \mathcal{ F}$
			\vspace{3mm}
			\item[3] $g_j(\tilde{\zeta}, \widetilde{\mathbf{W}}^m) =  \tilde{g}_j(\tilde{\zeta}, \widetilde{\mathbf{W}}^m; \widetilde{\mathbf{W}}^m) \quad \forall \hspace{2mm}\left\lbrace \tilde{\zeta}, \widetilde{\mathbf{W}}^m \right\rbrace \in \mathcal{ F}$	
			\item[4] $ \nabla_{\mathbf {W}}  g_j(\tilde{\zeta}, \widetilde{\mathbf{W}}^m) = \nabla_{\mathbf {W}} \tilde{g}_j(\tilde{\zeta}, \widetilde{\mathbf{W}}^m; \widetilde{\mathbf{W}}^m)$
		\end{itemize}  
		By following similar steps to \cite{Alaa}, we conclude that sequence generated by algorithm \ref{alg1} converges to a KKT solution of the non-convex optimization problem \eqref{NewA}.
		\bibliographystyle{IEEEtran}\bibliography{IEEEabrv,refs}
		 \vskip -2\baselineskip plus -1fil
	\begin{IEEEbiography}[{\includegraphics[width=1in,height=1.25in,clip,keepaspectratio]{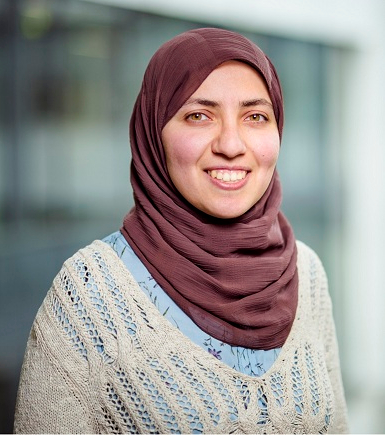}}]{Aya Mostafa Ahmed} received the B.Sc. and M.S.C
	degrees in electrical engineering from the German University in Cairo, Egypt, in 2011 and 2014 respectively.
	She is currently pursuing the Ph.D. degree with the Institute of Digital Communication Systems, Ruhr-Universität Bochum, Germany. Her research interests include MIMO radar signal processing, waveform design optimization, cognitive radars, direction of arrival (DOA) algorithms and machine learning applications for radar resources management. 
	\end{IEEEbiography}  \vskip -2\baselineskip plus -1fil
	\begin{IEEEbiography}[{\includegraphics[width=1in,height=1.25in,clip,keepaspectratio]{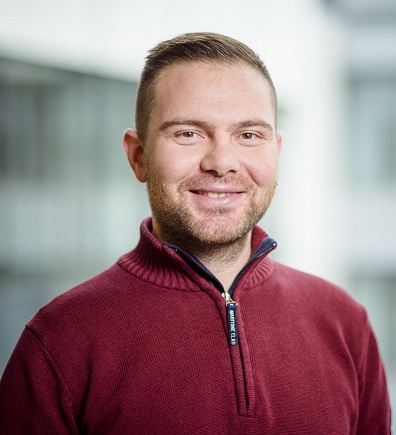}}]{Alaa Alameer Ahmad}
	received his B.Sc. degree in electrical engineering from Higher Institute of Applied Sciences and Technology (Hiast), Damascus, Syria in 2008, his M.Sc. degree in information and communication technology from TU-Darmstadt, Darmstadt, Germany, in 2015. He received the AKDN scholarship for the year 2011-2012. He is currently pursuing the Ph.D. degree with the Institute of Digital Communication Systems, Ruhr-University Bochum, Germany. His research interests include optimization of wireless communication systems, signal processing in communication and machine learning application for resources management. 
	\end{IEEEbiography} 
 \vskip -2\baselineskip plus -1fil
	\begin{IEEEbiography}[{\includegraphics[width=1in,height=1.25in,clip,keepaspectratio]{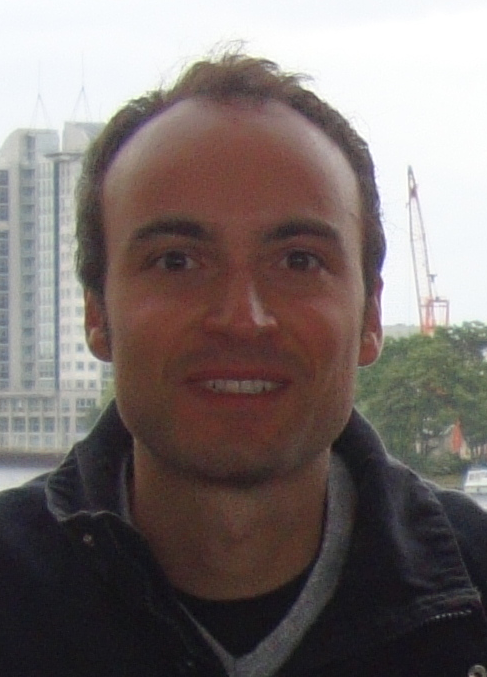}}]{Stefano Fortunati}
 received the graduate degree in telecommunication engineering and the Ph.D. degree, both from the University of Pisa, Italy, in 2008 and 2012, respectively. In 2012, he joined the Department of Ingegneria dell'Informazione, University of Pisa, where he was a researcher with a postdoc position until September 2019. Since October 2019, he is an associate researcher at Université Paris-Saclay, CNRS, CentraleSupélec, Laboratoire des signaux et systems (L2S), 91190, Gif-sur-Yvette, France. From Sept. 2020 he is a permanent lecturer (enseignant-chercheur) at IPSA in the Parisian campus of Ivry-sur-Seine. From September 2012 to November 2012 and from September 2013 to November 2013, he was a Visiting Researcher with the CMRE NATO Research Center, La Spezia, Italy. From May 2017 to April 2018, he spent a period of one year as a Visiting Researcher with the Signal Processing Group, Technische Universitat Darmstadt. He was a recipient of the 2019 EURASIP JASP Best Paper Award. His professional expertise encompasses different areas of the statistical signal processing, with particular focus on point estimation and hypothesis testing, performance bounds, misspecification theory, robust and semiparametric statistics and statistical learning theory.
	\end{IEEEbiography} 
	\begin{IEEEbiography}[{\includegraphics[width=1in,height=1.25in,clip,keepaspectratio]{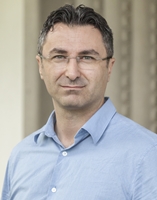}}]{Aydin Sezgin}
 received the Dipl.Ing. (M.S.) degree in communications engineering from Technische Fachhochschule Berlin (TFH), Berlin, in 2000, and the Dr. Ing. (Ph.D.) degree in electrical engineering from TU Berlin, in 2005.
From 2001 to 2006, he was with the Heinrich-Hertz-Institut, Berlin. From 2006 to 2008, he held a postdoctoral position, and was also a lecturer with the Information Systems Laboratory, Department of Electrical Engineering, Stanford University, Stanford, CA, USA. From 2008 to 2009, he held a postdoctoral position with the Department of Electrical Engineering and Computer Science, University of California, Irvine, CA, USA. From 2009 to 2011, he was the Head of the Emmy-Noether- Research Group on Wireless Networks, Ulm University. In 2011, he joined TU Darmstadt, Germany, as a professor. He is currently a professor of information systems and sciences with the Department of Electrical Engineering and Information Technology, Ruhr-Universität Bochum, Germany. He is interested in signal processing, communication, and information theory, with a focus on wireless networks. He has published several book chapters more than 50 journals and 170 conference papers in these topics. He has coauthored a book on multi-way communications. Aydin is a winner of the ITG-Sponsorship Award, in 2006. He was a first recipient of the prestigious Emmy-Noether Grant by the German Research Foundation in communication engineering, in 2009. He has coauthored papers that received the Best Poster Award at the IEEE Communication Theory Workshop, in 2011, the Best Paper Award at ICCSPA, in 2015, and the Best Paper Award at ICC, in 2019. He has served as an Associate Editor for the IEEE Transactions on Wireless Communications, from 2009 to 2014.
	\end{IEEEbiography} 
 \vskip -2\baselineskip plus -1fil
\begin{IEEEbiography}[{\includegraphics[width=1in,height=1.25in,clip,keepaspectratio]{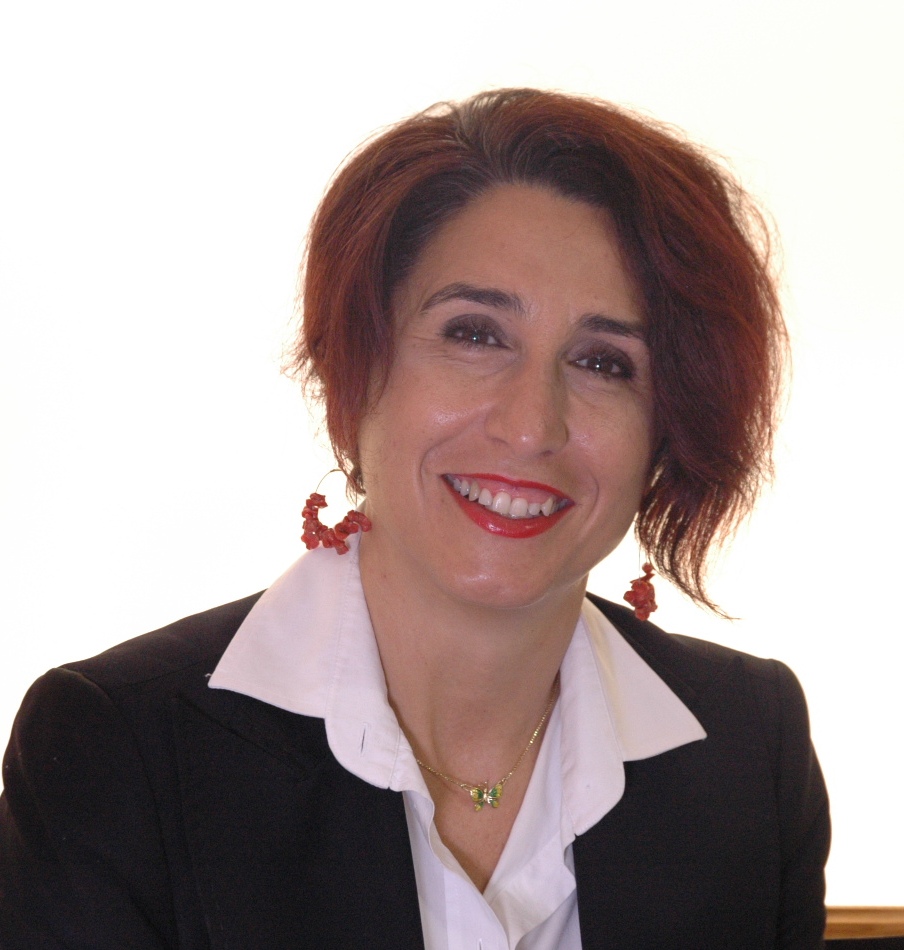}}]{Maria Sabrina Greco }
	graduated in Electronic Engineering in 1993 and received the Ph.D. degree in Telecommunication Engineering in 1998, from University of Pisa, Italy. From December 1997 to May 1998 she joined the Georgia Tech Research Institute, Atlanta, USA as a visiting research scholar where she carried on research activity in the field of radar detection in non-Gaussian background. 
	In 1993 she joined the Dept. of Information Engineering of the University of Pisa, where she is Full Professor since 2017. She’s IEEE fellow since Jan. 2011. She was co-recipient of the 2001 and 2012 IEEE Aerospace and Electronic Systems Society’s Barry Carlton Awards for Best Paper, co-recipient of 2019 EURASIP JASP Best Paper Award, and recipient of the 2008 Fred Nathanson Young Engineer of the Year award for contributions to signal processing, estimation, and detection theory and of IEEE AESS Board of Governors Exceptional Service Award for “Exemplary Service and Dedication and Professionalism, as EiC of the IEEE AES Magazine”. In May-June 2015 and in January-February 2018 she visited as invited Professor the Université Paris-Sud, CentraleSupélec, Paris, France. 
	She has been general-chair, technical program chair and organizing committee member of many international conferences over the last 10 years. She has been lead-guest editor for the special issue on “Advances in Radar Systems for Modern Civilian and Commercial Applications”, IEEE Signal Processing Magazine, July/September 2019, guest editor of the special issue on “Machine Learning for Cognition in Radio Communications and Radar” of the IEEE Journal on Special Topics of Signal Processing, lead guest editor of the special issue on "Advanced Signal Processing for Radar Applications" of the IEEE Journal on Special Topics of Signal Processing, guest co-editor of the special issue of the Journal of the IEEE Signal Processing Society on Special Topics in Signal Processing on "Adaptive Waveform Design for Agile Sensing and Communication," and lead guest editor of the special issue of International Journal of Navigation and Observation on” Modelling and Processing of Radar Signals for Earth Observation. She’s Associate Editor of IET Proceedings – Sonar, Radar and Navigation, and IET-Signal Processing, and Editor in Chief of the Springer Journal of Advances in Signal Processing (JASP). She’s member of the IEEE AESS Board of Governors and has been member of the IEEE SPS BoG (2015-17) and Chair of the IEEE AESS Radar Panel (2015-16). She has been as well SPS Distinguished Lecturer for the years 2014-2015, AESS Distinguished Lecturer for the years 2015-2020, and AESS VP Publications (2018-2020). She’s now IEEE SPS Governor-at-Large for Region 8.
	Her general interests are in the areas of statistical signal processing, estimation and detection theory. In particular, her research interests include clutter models, coherent and incoherent detection in non-Gaussian clutter, CFAR techniques, radar waveform diversity and bistatic/mustistatic active and passive radars, cognitive radars. She co-authored many book chapters and about 200 journal and conference papers.
\end{IEEEbiography} 
\begin{IEEEbiography}[{\includegraphics[width=1in,height=1.25in,clip,keepaspectratio]{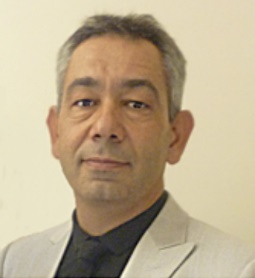}}]{Fulvio Gini }
(Fellow IEEE) received the Doctor Engineer (cum laude) and the Research Doctor degrees in electronic engineering from the University of Pisa, Italy, in 1990 and 1995 respectively. In 1993 he joined the Department of Ingegneria dell'Informazione of the University of Pisa, where he become Associate Professor in 2000 and he is Full Professor since 2006. Prof. Gini is the Deputy Head of the Department since November 2016. From July 1996 through January 1997, he was a visiting researcher at the Department of Electrical Engineering, University of Virginia, Charlottesville. He is an Associate Editor for the IEEE Transactions on Aerospace and Electronic Systems (AES) since Jan. 2007 and for the Elsevier Signal Processing journal since Dec. 2006. He has been AE for the Transactions on Signal Processing (2000–06) and is a Senior AE of the same Transaction since February 2016. He was a Member of the EURASIP JASP Editorial Board. He was co-founder and 1st co-Editor-in-Chief of the Hindawi International Journal on Navigation and Observation (2007-2011). He was the Area Editor for the Special issues of the IEEE Signal Processing Magazine (2012-14). He was co-recipient of the 2001 and 2012 IEEE AES Society's Barry Carlton Award for Best Paper published in the IEEE Transactions on AES and co-recipient of the 2020 EURASIP JASP Best Paper Award. He was recipient of the 2020 EURASIP Meritorious Service Award, of the 2003 IEE Achievement Award, and of the 2003 IEEE AES Society Nathanson Award to the Young Engineer of the Year. He is the IEEE AES Society Awards Chair. He is Member of the IEEE AES Society Radar System Panel (2008-present), member of the IEEE AES Society Board of Governors (BoG) (2017-2022) and member of the IEEE Signal Processing Society Board of Governors (2021-2023). He was a member of the IEEE SPS Awards Board (2016-2018) and he is presently a member of the IEEE SPS Conference Board (2019-2020). He has been a Member of the Signal Processing Theory and Methods (SPTM) Technical Committee (TC) of the IEEE Signal Processing Society and of the Sensor Array and Multichannel (SAM) TC for many years. He is a member of the IEEE TAB Awards and Recognition Committee (TABARC). He was a Member of the Board of Directors (BoD) of the EURASIP Society, the Award Chair (2006-2012) and the EURASIP President (2013-2016). He is the General co-Chair of the 2020 IEEE Radar Conference, Florence (Italy). He was the Technical co-Chair of the 2006 EURASIP Signal and Image Processing Conference (EUSIPCO 2006), Florence (Italy), of the 2008 Radar Conference, Rome (Italy), and of the 2015 IEEE CAMSAP workshop, Cancun (Mexico). He was the General co-Chair of the 2nd Workshop on Cognitive Information Processing (CIP2010), of the 2014 IEEE International Conference on Acoustics, Speech and Signal Processing (ICASSP 2014), and of the 2nd, 3rd and 4th editions of the workshop on Compressive Sensing in Radar (CoSeRa). He was the Section Editor for the "Radar Signal Processing" section, Vol. 3 of the Academic Press Library in Signal Processing, S. Theodoridis and R. Chellappa editors, Elsevier Ltd, 2013. He was the guest co-editor of two special sections of the Journal of the IEEE SP Society on Special Topics in Signal Processing, one on "Adaptive Waveform Design for Agile Sensing and Communication" (2007) and the other on "Advanced Signal Processing for Time/Frequency Modulated Arrays" (2017), guest editor of the special section of the IEEE Signal Processing Magazine on "Knowledge Based Systems for Adaptive Radar Detection, Tracking and Classification" (2006), guest co-editor of the two special issues of the EURASIP Signal Processing journal on "New trends and findings in antenna array processing for radar" (2004) and on "Advances in Sensor Array Processing (in memory of Alex Gershman)" (2013). He is co-editor and author of the book "Knowledge Based Radar Detection, Tracking and Classification" (2008) and of the book "Waveform Diversity and Design" (2012). He authored or co-authored 11 book chapters, 141 journal papers, and 172 conference papers. H-index: 47, 8446 citations (source: Google Scholar).
		\end{IEEEbiography} 
\end{document}